\documentclass[twocolumn]{aastex631}

  \makeatletter
  \long\def\ydeleted#1{\iftrack{\global\advance\refchangenumber by 1\relax%
  \vtop to 0pt{\vss
  \hypertarget{link\the\refchangenumber}{}
  \vskip14pt}%
  \ifnumlines\ifabstract\else%
  \xdef\doit{\noexpand\linelabel{\the\refchangenumber}}\doit\fi%
  \else%
  \xdef\doit{\noexpand\label{\the\refchangenumber}{}{}{}}\doit%
  \fi}%
  \begingroup\color{trackchange}\bfseries%
  \ifbib\let\sout\relax\fi%
  (Deleted: \sout{#1})\endgroup%
  \ifabstract\label{\the\refchangenumber}%
  \expandafter\gdef\csname changenum\the\refchangenumber\endcsname{Deleted:
  {\color{trackchange}\bfseries
  \sout{#1}}\global\silenttrue}%
  \else
  \expandafter\gdef\csname changenum\the\refchangenumber\endcsname{Deleted:
  {\color{trackchange}\bfseries%
  \sout{#1}}\global\silentfalse}\fi%
  \fi}
  \makeatother

\usepackage{amsmath}
\usepackage{amsfonts}
\usepackage{amssymb}

\shorttitle{MWR Observations of Giant Planet Atmospheres}
\shortauthors{Akins and Hofstadter}

\graphicspath{{./}{figures/}}

\begin{document}

\title{Microwave Remote Sensing of Giant Planet Atmospheres: Considerations for Saturn and Uranus Orbiter Missions}


\author[0000-0001-8379-1909]{Alex Akins}
\affiliation{Jet Propulsion Laboratory, California Institute of Technology, Pasadena, CA, USA}
\email{alexander.akins@jpl.nasa.gov}
\author[0000-0002-3208-3918]{Mark Hofstadter}
\affiliation{Jet Propulsion Laboratory, California Institute of Technology, Pasadena, CA, USA}
\affiliation{Space Science Institute, Boulder, CO, USA}

\begin{abstract}
Microwave radiometer instruments are uniquely capable of studying volatile element abundances and heat/mass transport within the deep atmospheres of giant planets. The results of the Juno MWR experiment have challenged long-held assumptions regarding Jupiter's troposphere, and the inclusion of such an instrument on future missions to Saturn or the ice giants would similarly re-write our understanding of giant planet atmospheres. We present a high-level summary of the capabilities of microwave radiometers for the study of giant planet atmospheres, including a review of the various scientific motivations and current knowledge, concise statement of the fundamentals of radiometer remote sensing accessible at the systems-engineering level, and calculations which are of illustrative value in the design of future instruments.
\end{abstract}


\section{Introduction} \label{sec:intro}

Microwave radiometer instruments (MWRs) are routinely flown on Earth science missions to study atmospheric and hydrological variables (e.g. soil moisture, ocean winds, atmospheric temperature). As such, a robust literature has emerged regarding their operation and calibration \citep{Ulaby}, and the resulting datasets have found widespread use in weather forecasting \citep{bormann_evaluation_2013, duncan_addition_2021}. Their applications towards the study of other planets, however, are less frequently discussed. This is almost certainly a result of the relative sparsity of deep space missions, but it is also a result of the significant differences in the applications of the measurement. These differences may hamper non-experts in developing an understanding of the capabilities and requirements of the instrument during the planetary science mission formulation process. One aim of this article is to provide a concise summary of microwave radiometer capabilities in the domain of giant planet atmospheric sounding to which the curious (but time-limited) researcher might refer. This is but one of the applications of this instrument in the exploration of giant planet systems (see e.g. \cite{Wolfenbarger2026} for a discussion of the applications for remote sensing of giant planet moons). 

Only two\footnote{Other heterodyne radiometers operating at millimeter wavelengths have also been fielded, such as the Rosetta MIRO and JUICE SWI instruments} multi-channel microwave radiometers have been intentionally flown on deep space missions: the Juno MWR \citep{Janssen2017, bolton_microwave_2021}, and the Mariner 2 radiometer \citep{barath_mariner_1963}. As the Mariner 2 instrument was the first spacecraft MWR ever fielded, its absolute and relative calibration was poor. It nevertheless achieved its primary function of measuring Venus' limb-darkening, thereby confirming that the planet's surface was warm \citep{Pollack1967}. By comparison, the Juno MWR built on years of intervening progress in radiometer technology. It has proven quite well-calibrated \citep{misra_calibration_2019} and obtained observations over a wide range of wavelengths and emission angles \citep{Janssen2017, Oyafuso2020, Zhang2020}, enabling detailed characterization of Jupiter's deep troposphere.  As the Juno mission approaches its conclusion, it is worthwhile to consider the successes of its MWR, particularly in view of the pre-mission performance assessments of an MWR-like instrument at Jupiter by \cite{Janssen2005} and \cite{DePater2005}. By assessing the differences in the state of knowledge regarding Jupiter's atmosphere before and after Juno MWR measurements, we aim also to demonstrate the value of this instrument for future missions that will study the more distant giant planets. This assessment is particularly relevant for the potential future NASA Uranus Orbiter and Probe flagship mission, as well as other missions which may visit the Saturn system. 

We begin our discussion by reviewing the scientific motivation for microwave remote sensing of giant planet atmospheres and contextualizing the significance of the Juno results. We also include a brief discussion of the current state of knowledge of the microwave properties of Saturn and the ice giants as inferred from ground-based measurements. We then present some simple relationships connecting the raw measurement of a radiometer system (power at the receiver output) with  atmospheric quantities of interest. This includes discussion of the idealized equilibrium cloud condensation model for giant planet atmospheres, for which a computer code is provided. We then discuss the implications of different instrument trades, including calculations of microwave spectra and information content for ``best-guess'' models of Saturn and Uranus' atmospheres. This discussion also includes an assessment of the capability of current and planned ground-based observatories. 

\section{MWR Science Motivation and Current Knowledge}

There are three general rationales for flying microwave sounders on missions to study giant planet atmospheres:

\begin{enumerate}
\item Measurement of the abundances of condensible gases in the troposphere.  
\item Assessment of how heat and volatile gases are transported vertically in the troposphere. 
\item Assessment of synoptic-scale spatial inhomogeneity and temporal variability from the troposphere to the interior
\end{enumerate} 

Each of these observables has implications for understanding the complex formation and thermal evolution sequences of the planets. MWR observations (at wavelengths longer than 10 cm) can also be sensitive to non-thermal processes (e.g. lightning and strong auroral activity). While it is not expected that such processes can be observed in this way on most planets, we will discuss them here, as they feature prominently among the results from the Juno mission.      

\subsection{Composition Measurements}
Descent probes are ideally suited for measurements of noble gas and isotopic compositions of giant planet atmospheres, as these quantities are generally homogeneous in the observable atmosphere (see e.g. \cite{mandt_2020, zahnle_2024}). By contrast, probes can provide skewed inferences of the abundances of volatile gases, which are altered by weather and can exhibit spatial variations. The most prominent example of this is Galileo's inference of sub-solar O/H at Jupiter from measurements of water abundance, which has been ascribed to the probe entering a dry hot spot in Jupiter's north equatorial belt \citep{Orton1998, showman_1998}. Remote sensing instruments which can assess the global distribution of volatile gases are therefore the preferred tools.  Microwave radiometry in particular is favored for measuring the abundances of N, S, and O via associated gases (NH$_3$, H$_2$S, H$_2$O) beneath their respective condensational clouds; such measurements are prohibitive for instruments operating at shorter wavelengths. Observations of CO in the stratosphere or upper troposphere can provide an additional constraint on O/H, but only if the effects of exogenous contamination can be quantified and there is knowledge of the extent of vertical mixing in the deep troposphere \citep{Cavalie2024, hyder_supersolar_2025, Yang2026}. 

The abundance ratios of N/H, S/H, and O/H relative to the current solar values \citep{asplund_chemical_2021, lodders_solar_2025} in giant planet atmospheres provide clues as to where the planet formed and through what mechanisms volatiles were delivered during formation. Conceptually, giant planet composition should be governed primarily by formation location, the composition of the proximal nebula, and the gas-to-solid ratios in the nebula at that location. The latter two quantities are generally dependent on the proximity of the planet to different ice lines, such as those of (progressing outward from the proto-sun) H$_2$O,  NH$_3$, CO$_2$, H$_2$S, CH$_4$, CO, and N$_2$ \citep{mousis_role_2020}.      There are, however, many pieces of the formation process which further affect the final heavy element abundances of formed planets, including but not limited to the size distribution of accreted solids, timescales of the various accretion stages, and the occurrence (or non-occurrence) of migration; we refer interested readers to the review articles of \cite{Helled2014, Guillot2023}. 

An outstanding question prior to Juno was how it was possible to produce the apparent consistency of elemental enrichment ($\approx 3 \times$ solar) of both volatile and stable elements in Jupiter's atmosphere \citep{lunine_origin_2004}. A goal of the Juno MWR experiment was to confirm this value for N/H and to determine if O/H followed a similar trend; this value was confirmed in both cases \citep{Li2020, Li2024, Moeckel2023}. This is incompatible with the formation hypothesis of \cite{gautier_enrichments_2001} and \cite{Hersant2004} that clathration of volatiles within crystalline ice during the cooling of the protosolar nebula was the primary way of incorporating volatiles other than oxygen into Jupiter. In this case, O/H would have been closer to $\approx 10 \times$ solar. While this model is ruled out, several models have been formulated which are consistent with our revised knowledge of Jupiter's atmospheric composition and can explain them using different processes. In this context, accurate inferences of N/H, S/H, and O/H for Saturn from MWR measurements would be quite useful, as Saturn and Jupiter are more likely to be influenced by similar processes. \cite{oberg_jupiters_2019, Bosman2019} posit that Jupiter's core formed beyond the N$_2$ snowline in the outer nebula. This model places O/H towards the higher end of the Juno constraints and predicts constant enrichment across elements for Saturn. Several models (e.g. \cite{booth_chemical_2017, mousis_jupiters_2019, schneider_how_2021, aguichine_possible_2022}) assess the capacity of inward dust and pebble drift to enhance enrichment of heavy elements in the nebula during formation. These models do not require a more distant formation for Jupiter. \cite{schneider_how_2021} predicts N/H $<$ C/H $<$ S/H $\approx$ O/H for Saturn, with the difference between N/H and C/H dependent on the local abundance of small carbon grains. \cite{booth_chemical_2017} similarly predicts C/O $\geq$ 1 for Jupiter and $<$ 1 for Saturn. \cite{ohno_jupiters_2021} propose a model whereby accumulated dust at ice lines results in the formation of localized cold, shadowed regions. This cold region would freeze nebular gas closer to the proto-sun than their usual icelines. This model applied to Saturn predicts N/H $<$ C/H $\approx$ S/H $\approx$ O/H. 

Uranus and Neptune's  formation sequences have resulted in final states that is markedly different from Jupiter and Saturn.  The ice giants contain a smaller proportion of H\textsubscript2 and He relative to heavy elements than the gas giants, but a larger proportion than would be expected for planets forming at large distances from the Sun. This implies that nebular gas accretion was beginning just as the disk began to dissipate, preventing the more vigorous runaway gas accretion stage associated with Jupiter and Saturn \citep{Helled2020}. \cite{lambrechts_separating_2014} suggest that Uranus and Neptune's distance at formation prevented them from reaching the ``pebble isolation mass'' wherein the core's gravitational perturbation of the disk traps incoming pebbles and accelerates the process of nebular gas accretion. Since the planets are continuously preferentially accreting solids, their resulting heavy element abundances are enriched. Other processes which could increase heavy element enrichment include migration during formation \citep{shibata_origin_2020, Turrini2021} or post-formation enrichment via giant impacts \citep{valletta_possible_2022}. Relative to Jupiter and Saturn, there are generally fewer formation models that offer specific predictions of inter-species enrichment trends for Uranus and Neptune. In one example, \cite{mousis_insights_2024} investigated the effects of the compositions of the pebble solids considered by \cite{lambrechts_separating_2014}, finding that clathrated solids would exhibit relative homogeneity across N/H, S/H, and O/H enrichment. Future studies making predictions for the ice giants that could be tested with MWR measurements are needed. 

In their review, \cite{Guillot2023} accurately assert that ``in order to link the atmospheric composition with the formation process, it is critical to properly model the planetary evolution.'' In light of results from the Juno and Cassini missions, it is no longer reasonable to confidently equate inferred atmospheric composition on its own to a particular formation sequence. Prior to Juno, Jupiter was considered the ideal object in terms of inferring bulk composition from atmospheric measurements, as its high internal heat flux was thought to ensure thorough mixing and relative compositional homogeneity through the envelope \citep{guillot_interior_2004}. Instead, Juno revealed that even Jupiter exhibits significant heterogeneity in its interior \citep{helled_revelations_2022}. In order to use MWR-derived atmospheric composition as a constraint on the formation sequence of giant planets, we must also obtain knowledge of the current extent of interior stratification (via gravity and seismology measurements) and models of how the interior has evolved over time (e.g. \cite{Muller2024}). Future observations will be aided by the advancement of comprehensive giant planet evolution codes (e.g. \cite{sur_evolution_2025, Helled2025}) and their application towards the forward modeling of present-day volatile abundances in giant planet atmospheres.

\subsection{Heat Transport and Atmospheric Dynamics} 

 Beyond measurement of the abundances of condensible gases at depth,  MWR observations of the vertically- and horizontally-resolved distribution of condensible gases, particularly their departure from simple equilibrium models, can provide insight into the processes by which heat and mass are transported in the troposphere of giant planets. MWR measurements are also sensitive to the thermal structure of the deep tropospheres of the giant planets below the 1-bar level which is commonly used to anchor the boundary of interior models. As we discuss further below, equilibrium models of giant planet atmospheric composition and temperature assume a constant abyssal trace gas composition which is affected solely by the condensation of the various cloud layers (in order and starting at greatest depth, H$_2$O, NH$_4$SH, NH$_3$, H$_2$S). One of the important successes of Juno MWR was the inference of just how substantially Jupiter's atmosphere deviates from these models.  This has opened a new window into the study of dynamics and transport in the atmosphere, which was one of the expectations prior to Juno's launch \citep{Janssen2005}.

Prior to Juno's arrival at Jupiter, models based on local equilibrium predicted the abundance of NH$_3$, the primary microwave absorber in the atmosphere, would vary strongly in and above the associated cloud near 0.7 bars due to condensation and atmospheric circulation, but the abundance would be relatively uniform horizontally and vertically below that cloud.  While Juno did find the most intense variations (order-of-magnitude) associated with the NH$_3$ cloud altitude, it also revealed factor-of-two variations at greater depths \citep{Li2017, Fletcher2021,bolton_microwave_2021,Moeckel2023}. It appears that dynamical processes are more efficient than previously thought at moving mass in the atmosphere and maintaining these gradients, via horizontally- and vertically-alternating circulation cells \citep{Fletcher2021,Duer2021}, and/or via relatively intense vertical motions . \cite{Guillot2020a, Guillot2020b} hypothesize that a steady stream of frozen water particles lift from $\sim$10 bars up to $\sim$0.7 bar, absorb NH$_3$ vapor in a ``mushball'' assemblage, and then fall to ~100 bars, resulting in strong evaporative downdrafts. Future MWR observations of the ice giants could determine if similar processes occur in their tropospheres.     . Juno data have also revealed that most storms and vortices in Jupiter's atmosphere do not extend deeper than $\sim$10 bars \citep{Fletcher2020, brueshaber2025, gavriel_dynamical_2025}, while stronger features, such as the Great Red Spot and high latitude energetic eddy systems, extend deeper \citep{bolton_microwave_2021, fletcher_structure_2026} and can significantly perturb atmospheric composition \citep{moeckel_tempests_2025}. 

A particularly important result from Juno MWR observations is the confirmation that convection-stable regions with super-adiabatic temperature gradients are present in Jupiter's troposphere \citep{Li2024}. Depending on the extent of atmospheric volatile enrichment (and thereby water, methane available to condense), stable regions like these can either be transient, periodically interrupted, or persistent. \cite{li_moist_2015, Li2023} hypothesize that Saturn's massive storms are the result of a decadal cycle of convective instability, re-establishment of stability via condensation, and radiative cooling in the troposphere. \cite{Friedson2017, Markham2021} find that, if persistent stable regions are present in the methane and water (and perhaps hydrogen sulfide, \cite{Ge2024}) clouds of the ice giants, primordial heat may remain trapped in the deeper envelope and interior.

 \subsection{Existing Constraints for Saturn and the Ice Giants} 

Observational constraints on the deep tropospheres of Saturn and the ice giants exist solely from ground-based measurements \citep{DePater2023} and from 2 cm radiometry of Saturn with Cassini \citep{Laraia2013}. Saturn's disk-integrated spectrum can be explained with $\approx 3 \times$ solar N/H and $> 5 \times$ solar S/H. The constraint of elevated S/H derived by \cite{Briggs1989}, however, appears driven by the need to reduce the abundance of NH$_3$ above the NH$_4$SH condensation altitude in the ECCM. Another possible explanation could be that Saturn's giant storms dry out the cloud-level atmosphere faster than the general circulation can resupply NH$_3$ vapor to saturation levels \citep{Li2023}, and that N/H and S/H are consistent. Uranus and Neptune, however, show clear evidence of preferential enrichment in S/H over N/H at equatorial latitudes \citep{dePater1991}, which can be concluded from the simple spectroscopic detection of H$_2$S gas above the NH$_4$SH cloud deck near the tropopause \citep{Irwin2018,Irwin2019}. Ground-based observations suggest S/H in excess of $30 \times$ solar and N/H nearer to solar for both planets at latitudes $<$ 45 degrees \citep{Molter2021, Tollefson2021}. However, at their poles, the abundances of H$_2$S and NH$_3$ must be quite close to explain the short-wavelength spectrum using equilibrium models \citep{Tollefson2021, Akins2023}. This has been explained as being a feature of the atmospheric circulation, with a deep step in composition restoring the relative abundances to the equatorial distribution. It is unclear if such a massive compositional gradient is consistent with the circulation strengths of the ice giants, and longer-wavelength resolved observations are necessary to obtain a firm grasp on the relative S/H and N/H values. Due to the lower temperatures, H$_2$O condensation occurs at deeper altitudes, which has thus far precluded firm estimates of O/H for Saturn or the ice giants from ground-based observations. Meter-wavelength radio telescope observations of Saturn are generally consistent with enrichments of $\geq 3 \times$ solar, whereas observations of Uranus have only recently become possible (see Appendix \ref{app:A}). Thermochemical calculations for Uranus and Neptune suggest upper limits between 30 and 300$\times$ \citep{Cavalie2024}, and the analysis of \cite{li_moist_2015} suggests that Saturn's storm frequency requires that O/H be elevated (order 10 $\times$ solar). We reiterate that some of the new dynamical processes being invoked to explain Juno MWR observations at Jupiter may also be active on Uranus and Neptune. Vertical and horizontal variations in opacity between the NH$_3$/H$_2$S clouds are several times larger at Uranus relative to Jupiter (e.g. \cite{Hofstadter2003}), so hypotheses which might explain this behavior consistently across planets are particularly useful.

\subsection{Non-thermal Processes and Ionized Media}
Due to the extremes of the Jupiter system, non-thermal sources of radiation can be observed at relatively short wavelengths. At wavelengths greater than 10 cm, Jupiter's magnetospheric synchrotron radiation is strong enough to obscure atmospheric thermal emission \citep{Bolton2004}; this was indeed one of the motivations for Juno to make microwave measurements from within the radiation belts. Beyond the radiation belts, signatures of auroral electron precipitation into the stratosphere have been observed in 20-50 cm MWR observations. \cite{bhattacharya_2025} found that penetration of auroral electrons into the lower stratosphere was capable of inducing microwave absorption strong enough to depress Juno MWR brightness temperatures, resulting in an unexpected new observable for the study of jovian auroral dynamics. From within the deeper troposphere, transient lightning signals have been observed from within Jupiter's water clouds. \cite{Brown2018, fletcher_structure_2026} characterized the distribution of these signals from Juno perijove passes and found more frequent occurrences north of 40$^\circ$N, and \cite{aglyamov2021} used these observations to constrain a model of lightning generation, providing an alternative inference on water abundance at the cloud base. Finally, Juno MWR observations have been used to infer atmospheric alkali metal ratios from the opacity incurred by Na and K thermal ionization in the deep troposphere of Jupiter \citep{bhattacharya_2023, Aglyamov2025}. This process has been hypothesized to explain the fact that 50 cm brightness temperatures measured by Juno MWR were globally colder than expected from NH$_3$ and H$_2$O opacity alone. It was initially suspected that this deviation could result from uncertainties in the opacity of water, but this was ruled out by laboratory experiments \citep{Steffes2023}. \cite{bhattacharya_2023, Aglyamov2025} found that sub-solar values of ionized Na/K could explain the 50 cm observations. This inference is surprisingly robust; an alternative hypothesis for low 600 MHz temperatures includes the presence of a deep atmosphere radiative zone, which itself would require low alkali metal abundances \citep{guillot_interior_2004}. In most cases, the ability of MWR measurements to study these processes came as a surprise. Unfortunately, it appears likely that these processes are most observable at Jupiter. Given current constraints on the radiation environment, synchrotron emission will likely not be observable at Saturn or the ice giants at wavelengths shorter than 10 meters (e.g. \cite{luthey1973}), and in a similar sense, auroral ions will likely not penetrate deeply enough in the collisional atmosphere to induce substantial microwave absorption. Of all of these processes, we suspect that lightning signals are the most likely to be detected, but this depends on the strength and frequency of the strokes. One shortcoming of Juno MWR for lightning detection was that the most rapid integration period was 100 ms, which is much longer than the expected duration of a sferic signal \citep{fletcher_structure_2026}; future MWR instruments intending to detect lightning should include a fast-integration mode.

\section{Modeling MWR Measurements}

\subsection{Radiometer System Basics} 
Microwave radiometers (defined here as operating between 3 m and 3 mm wavelength) are similar to mid- and far-infrared radiometers in that they measure primarily thermal emission, rather than reflected light, from the neutral atmospheres of giant planets. Calibrated measurements are similarly often reported as brightness temperatures in Kelvin units, and observation resolution, in the absence of noise, is diffraction-limited (i.e. controlled by the effective diameter of the telescope being used). One distinguishing characteristic for microwave radiometry is the validity of the Rayleigh Jeans approximation for radiance intensity $B_\nu$, specifically that thermal emission intensity is linearly proportional to blackbody brightness temperature $T_B$. This linear relationship allows the power $P_r$ received by a radiometer system with gain $G$ over a finite bandwidth $B$ to be expressed as an equivalent noise temperature $T_r$ such that $P_r=Gk_bT_rB$, where k$_b$ is the Boltzmann constant. This greatly simplifies the discussion of signal and noise in microwave receiver systems. 

The first intermediate quantity in the conversion between the scene brightness temperature $T_B$ (what we want to measure), and what is actually recorded by a radiometer is the ``antenna temperature'', which is the integral of the scene brightness temperature $T_B$ over 4$\pi$ steradians weighted by the antenna power pattern $A(\Omega)$. 
\begin{equation} \label{eq:T_A}
T_A = \frac{\iint A(\Omega) T_B d\Omega}{\iint A(\Omega) d\Omega}
\end{equation}
The power pattern $A(\Omega)$ of an antenna is proportional to the square of Fourier transform of the aperture field. For a uniformly illuminated square antenna aperture of dimension $l$, the power pattern in boresight-centered elevation $\theta$ and azimuth $\phi$ coordinates evaluates as
\begin{equation} 
A(\theta, \phi) \propto \left[\text{sinc}\left(\frac{l}{\lambda}\sin\theta\cos\phi\right) \text{sinc}\left(\frac{l}{\lambda}\sin\theta\sin\phi\right)\right]^2
\end{equation}

where $\text{sinc}(x) = \sin (\pi x)/ \pi x$. When modeling the antenna temperature for a particular scene, it is important to also take into account the antenna pattern side-/backlobes (i.e. all points for which $A(\Omega) < 1/2 \times A_\text{max}$) as well as antenna self-emission. The antenna temperature can be expressed as a sum of main beam, sidelobe, and self-emission terms
\begin{equation} 
T_A = \epsilon \eta T_B + \epsilon (1-\eta) T_S + (1-\epsilon) T_0,
\end{equation} 
for main beam efficiency $\eta$, radiation efficiency $\epsilon$, cold space background temperature $T_S$, and antenna physical temperature $T_0$. The main beam efficiency $\eta$ can be computed from $A(\Omega)$ by determining the fraction of the pattern which falls within the main beam full width at half maximum. The solid angle distribution of $T_S$ between emission from the planet and cold space depends on the details of the observing geometry. In most cases, the cold space background temperature can be specified as a uniform 2.725 K \citep{fixsen_temperature_2009}, although for wavelengths longer than 10 cm, the structure of galactic emission starts to become relevant \citep{Dinnat2015, anderson_radio_2025}. 

The received signal at the antenna is transmitted to the radiometer front-end, where it undergoes amplification and signal conditioning. In most cases, the band-limited signal is converted to a lower frequency baseband using a coherent mixer, whereas some designs for low frequency observations may omit this step. For analog receiver systems, the power at the output terminals of the radiometer front-end circuit is inferred by sampling the voltage across a detector diode. This voltage is converted to a pulse train using a voltage-to-frequency converter, and the number of pulses that occur within a finite integration period $\tau$ is recorded. This quantity, referred to as ``radiometer count'' number $C$, serves as the raw measurement from which $T_A$ (and eventually scene $T_B$) can be reconstructed. Next-generation Digital radiometers are capable of computing $C$ by processing the directly digitized baseband signal as opposed to using detector diodes. 

The count number when the receiver circuit is connected to the antenna, $C_A$ is linearly proportional to the antenna temperature $T_A$ as $C_A = G_s(T_A + T_E)$, where $T_E$ is the receiver effective noise temperature and $G_s$ is the system gain (note that we distinguish this from $G$ by factoring in the $k_b$ and $B$ terms in the previous definition). The system gain is the product of gains of all circuit components in series, and the receiver noise temperature is the sum of the noise temperatures of each component weighted by the gains of the elements which precede them (e.g. the noise temperature of an N-element device is $T_{E,1} + T_{E,2}/G_1 + T_{E,3}/(G_1G_2) $..., and so on). Common values of $T_E$ for microwave radiometers range from 100 to 1000 K, increasing further as wavelength decreases. For an ideal, linear radiometer system, antenna temperature can then be reconstructed as 

\begin{equation} \label{eq:rad_counts}
T_A = \frac{C_A}{G_s} - T_E
\end{equation}

While single-point calibration is possible (when $T_E$ has good stability and well-characterized physical temperature dependence, \cite{brown_2023}), the $G_s$ and $T_E$ terms are inherently time-variable, and observations of at least two calibration sources are common. Externally-calibrated radiometers use cold-target (e.g. space) and hot-target (e.g. an absorber panel on the spacecraft body) measurements to calibrate $T_A$. In this arrangement, the reconstruction relationship becomes 

\begin{equation} \label{eq:count}
T_A=T_\text{cold} + \frac{C_A-C_\text{cold}}{C_\text{hot}-C_\text{cold}}(T_\text{hot}-T_\text{cold} )
\end{equation}

where $T_\text{cold}$ and $T_\text{hot}$ are the brightness temperatures of the cold and hot targets, and $C_\text{cold}$ and $C_\text{hot}$ are the respective counts when measuring the cold and hot targets.  In Equation \ref{eq:count}, all $C$ terms are observed quantities and all $T$ term are either known or independently-measured parameters of the calibration sources, allowing $T_A$ to be determined. Internally-calibrated radiometers use switched observations of internal noise sources, including temperature-controlled matched impedance loads and noise diodes, to calibrate $T_A$. For example, an internally calibrated radiometer may use a reference load to calculate $T_A$ as 

\begin{equation} \label{eq:rad_counts}
T_A = \frac{C_A-C_\text{ref}}{G_s} + T_\text{ref} 
\end{equation}

and use the noise diodes to solve for $G_s$ as 

\begin{equation} 
G_s = \frac{C_\text{ND+ref} - C_\text{ref}}{T_{ND}}
\end{equation} 

The precision with which T$_A$ can be recovered is related to the bandwidth and the receiver integration time $\tau$ as 

\begin{equation} \label{eq:rase}
\Delta T_A = \frac{T_A + T_E}{\sqrt{B \tau}}
\end{equation}

Equation \ref{eq:rase} is the ideal limit of achievable precision for real aperture radiometers. An expanded (but still simplified) expression for $\Delta T_A$ is provided below which accounts for observations of the antenna port and a calibration reference source with some cycle fraction $F$ as well as the fluctuations in system gain over time. 

\begin{equation}
\begin{split} 
\Delta T_{A} = \biggl[ \frac{F(T_A + T_E)^2 + (1-F)(T_\text{ref} + T_E)^2}{B \tau} + \\ 
\left(\frac{\Delta G}{G}\right)^2 (T_A - T_\text{ref})^2 \biggr]^{1/2}
\end{split} 
\end{equation}

Readers interested in a more thorough review of the concepts introduced here can refer to the text by \cite{Ulaby} on microwave remote sensing. 

\subsection{Modeling Microwave Brightness Temperatures} 
  As a consequence of the validity of the Rayleigh-Jeans relationship, the familiar equations of radiative transfer can be modified to incorporate brightness temperature directly, rather than converting first to radiance. For the case of nadir sounding with minimal reflection of radiation from space (generally appropriate for sounding abyssal giant planet atmospheres), the brightness temperature observed by a downward-looking radiometer can be computed by integrating the product of the atmospheric temperature $T(s)$ and the frequency-dependent temperature weighting function $W(f,s)$ along the line-of-sight path $s$ (defined as zero at the location of the observer and increasing downward towards the planet). 
\begin{equation} \label{eq:rt}
T_B = \int_{\infty}^0 W(f, s) T(s) ds
\end{equation}
\begin{equation} 
W(f, s) = \kappa(f, s)e^{-\tau(f, s)} 
\end{equation} 
\begin{equation} 
\tau(f, s) = \int_s^0 \kappa(f, s')ds' 
\end{equation} 
Here, $\kappa$ represents the atmospheric absorption coefficient (defined as absorption per unit path length), and $\tau$ is the integrated optical depth (dimensionless). For shallow observing angles, it is common to approximate the path $s$ as the planet-centered vertical atmospheric coordinate $r$ divided by the cosine of the radiation emission angle i.e. $s = r_\text{obs}-r \sec(\theta)$, but for microwave atmospheric sounding, it is not uncommon for the optical path to be subject to significant refractive bending. In this case, the path coordinate can be computed from Snell's law and the Abel integral transform \citep{Thompson1982}. For $a=nr\sin(\theta)$, computed at the top of the atmosphere, the path $s$ can be computed from 
\begin{equation} 
s(r) = r_\text{obs}-\int_r^{r_\text{obs}} \frac{n(r')r'}{\sqrt{(n(r')r')^2 - a^2}}dr' 
\end{equation}
While Equation \ref{eq:rt} could be elaborated to include the effects of cloud scattering, currently available constraints on cloud particle size distributions (likely small) and composition (mostly ice) render negligible scattering cross-sections for notional giant planet aerosols, except potentially in the case of the liquid water cloud. Since the opacity of gas mixtures in the microwave is generally lower than in the infrared, microwave observations are sensitive to thermal emission from deeper in the tropospheres of the giant planets. The shapes of giant planet microwave spectra result from highly pressure-broadened rotation lines \citep{Townes1955} for polar molecules, as opposed to rotation-vibration spectrum, which is more relevant in the far-infrared. The spectra of non-polar gases (e.g. H\textsubscript2, He, CH\textsubscript4) results instead from collision-induced absorption and/or quadrupole-quadrupole interactions. Laboratory measurements of the microwave absorption spectrum for abundant trace gases relevant to giant planet microwave sounding (specifically NH$_3$, H$_2$S, H$_2$O, and PH$_3$ broadened by high pressure H$_2$, He and CH$_4$ mixtures)  were obtained in the lead-up to the Juno mission, and as a result, good models for $\kappa$ are available in the literature \citep{Steffes2017,Steffes2023}. Readers who are interested in a more thorough review of microwave radiative transfer concepts introduced here can refer to the text by \cite{Janssen1993} on atmospheric microwave sounding. 

Models of giant planet atmospheric composition and temperature are needed to compute the upwelling brightness temperature from Equation \ref{eq:rt}. Equilibrium cloud condensation models (ECCM's) can be applied for this purpose, wherein the deepest atmosphere is assumed to be well-mixed. The vertical profiles of condensible gases are determined by pre-specified trace gas deep abundance and the saturation vapor pressure rules for aerosol condensation. Although their status as state-of-the-art for 1D modeling has been superseded by cloud-resolving models \citep{Sugiyama2014, Li2019, Ge2024}, they remain useful tools for simulating radiometric observations due to their relative computational simplicity. 

Since the ECCM we describe here does not include photochemistry or radiative heating and cooling, which would cause the atmospheric lapse rate to deviate from an adiabat near the radiative-convective boundary, the first step taken is to provide an input temperature profile, e.g. from a radio occultation dataset \citep{Schinder2011, Lindal1987}, which is extrapolated to the desired depth using the dry adiabatic lapse rate. 

\begin{equation} 
\frac{d\ln T}{d \ln p} = \frac{R}{c_p}
\end{equation}

Here, $R$ and $c_p$ are the gas constant and isobaric heat capacity. These quantities can be calculated either assuming ideal gases or, for higher pressures, using equations of state for non-ideal gases (our calculations use the equation of state of \cite{karpowicz2013} for H\textsubscript2/He/CH\textsubscript4/H\textsubscript2O mixtures). Values of the deep atmosphere mole fraction  $x$ for H\textsubscript{2}O, NH\textsubscript{3}, H\textsubscript{2}S, and CH\textsubscript{4} are then specified, as well as the bulk H\textsubscript{2}/He abundances. Model construction proceeds iteratively upwards; at each step, the dry lapse rate is calculated, and the saturation pressures of the gases are checked. When the trace gases exceed their respective saturation vapor pressures, condensation occurs. 





The lapse rate and the change in mole fraction for each condensing constituent are then determined per \cite{Li2018} as

\begin{equation} 
\frac{d\ln T}{d\ln P} = 
\frac{1 + 
\sum_i \frac{L_in_i}{RT}
}
{\frac{c_p}{R} 
+ \frac{\sum_{i} \frac{L_i^2n_i}{(RT)^2} 
+ \left(\sum_i \frac{L_in_i}{RT}\right)^2}{1 + \sum_i n_i}
}
\end{equation}

\begin{equation} \label{eq:mf}
d\ln x_i = \frac{L_i}{RT}d\ln T - d\ln p + \sum_{j\neq i} n_j  \left(\frac{L_j}{RT}d\ln T - d\ln p \right) 
\end{equation}

Here, $n$ is the gas volume mixing ratio with respect to dry air. In the limit of small $n$, these equations reduce to 

\begin{equation} 
\frac{d\ln T}{d\ln P} = \frac{1 + \sum_i \frac{L_i x_i}{RT}}{\frac{c_p}{R} + \sum_i \frac{L_i^2x_i}{(RT)^2}}
\end{equation}

\begin{equation} \label{eq:small_mf}
d\ln x = \frac{L_i}{RT}d\ln T - d\ln p 
\end{equation}

The lapse rates and condensation rules for NH$_4$SH are based on its equilibrium constant rule, which modifies the reduced equations as 

\begin{equation} 
\frac{d\ln T}{d\ln P} = \frac{1 + \frac{2L}{RT}\left(\frac{x_\mathrm{NH_3} x_\mathrm{H_2S}}{x_\mathrm{NH_3}+x_\mathrm{H_2S}}\right)}{\frac{c_p}{R} +  \
\frac{10834L}{RT^2}\left(\frac{x_\mathrm{NH_3}x_\mathrm{H_2S}}{x_\mathrm{NH_3}+x_\mathrm{H_2S}}\right)}
\end{equation}

\begin{equation} 
	dx_\mathrm{NH_3} = dx_\mathrm{H_2S} = \frac{x_\mathrm{NH_3}x_\mathrm{H_2S}}{x_\mathrm{NH_3} + x_\mathrm{H_2S}} \left(\frac{10834d \ln T}{T} - 2d \ln P\right)
\end{equation}
Condensation of the H$_2$O solution cloud is carried out by checking at each step for the NH$_3$ solution concentration which zeros the following equation using a root-finding algorithm
\begin{equation}
f(C) = (1-C) (p_\mathrm{NH_3, sat} - p_\mathrm{NH_3}) - C(p_\mathrm{H_2O, sat} - p_\mathrm{H_2O}) 
\end{equation}
The resulting value of $f(C)$ is compared to the boundary values f(0) and f(1). If one of the boundary values achieves a lesser $f(C)$, then the condensation of the corresponding pure component is checked instead of the solution condensation. If the solution condenses, then the change in $X_{H_2O}$ is determined via Equation \ref{eq:mf}, the change in $x_{NH3}$ is determined as 
\begin{equation} 
dx_\mathrm{NH_3} = \left(\frac{C}{1-C}\right) dx_\mathrm{H_2O}, 
\end{equation} 
\cite{atreya_photochemistry} provide the relevant expression for determining the corresponding dissolution of H$_2$S. 

The density of the cloud aerosols resulting from condensation at each discrete step is computed as 

\begin{equation} 
D = \frac{m p^2 \Delta x}{RT \Delta p}
\end{equation} 

The expression above represents an upper limit on possible cloud densities, with the true value likely lower (see e.g. \cite{Ackerman2001, Wong2015}). Prior analyses of giant planet microwave observations have neglected opacities due to ice clouds, which are expected to be low. The effects of the liquid water cloud, however, should be further assessed in the context of parameter retrieval studies (e.g. \cite{Moeckel2023}). 

This iterative computation of temperature, gas mole fraction, and cloud aerosol profiles proceeds upwards until the lower bound of the input occultation profile is reached. The entire procedure is then run again with an adjustment to the starting deep temperature so that the final temperature profile extrapolates smoothly from the input. Finally, the relationship between pressure and altitude ($r$, distance from the body center) is calculated on different grid points $i$ using the hypsometric equation for an air mixture with a mean molar mass $M$. 

\begin{equation} 
r_{i+1} = r_i + \frac{RT}{Mg}\ln\left(\frac{p_{i}}{p_{i+1}}\right)
\end{equation}

We neglect here the differences between the specific heats for different ortho-/para-hydrogen fractions, and incorporation of the zonal gravity field into the final pressure/altitude profile. Further details regarding ECCM construction are available in the original papers by \cite{Lewis1969, Weidenschilling1973} and in applications to analyses of giant planet microwave spectra by \cite{DePater1985,Briggs1989, DeBoer1995}. The interested reader is also referred to \cite{Li2018} for a more contemporary and complete discussion of the interplay between lapse rate and gas condensation in giant planet atmospheres.

\begin{figure}[htbp]
   \centering
   \includegraphics[width=\linewidth]{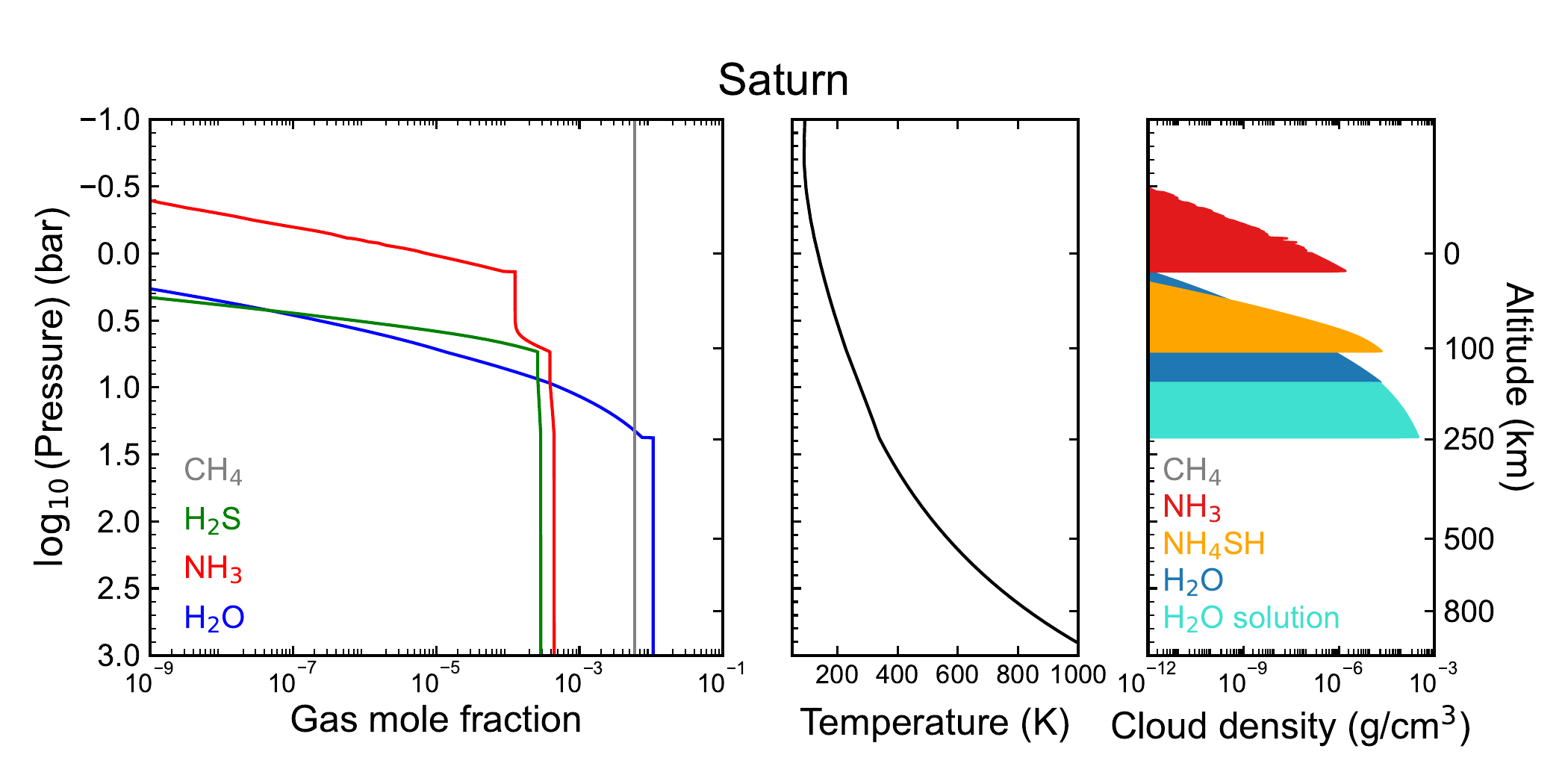} 
   \includegraphics[width=\linewidth]{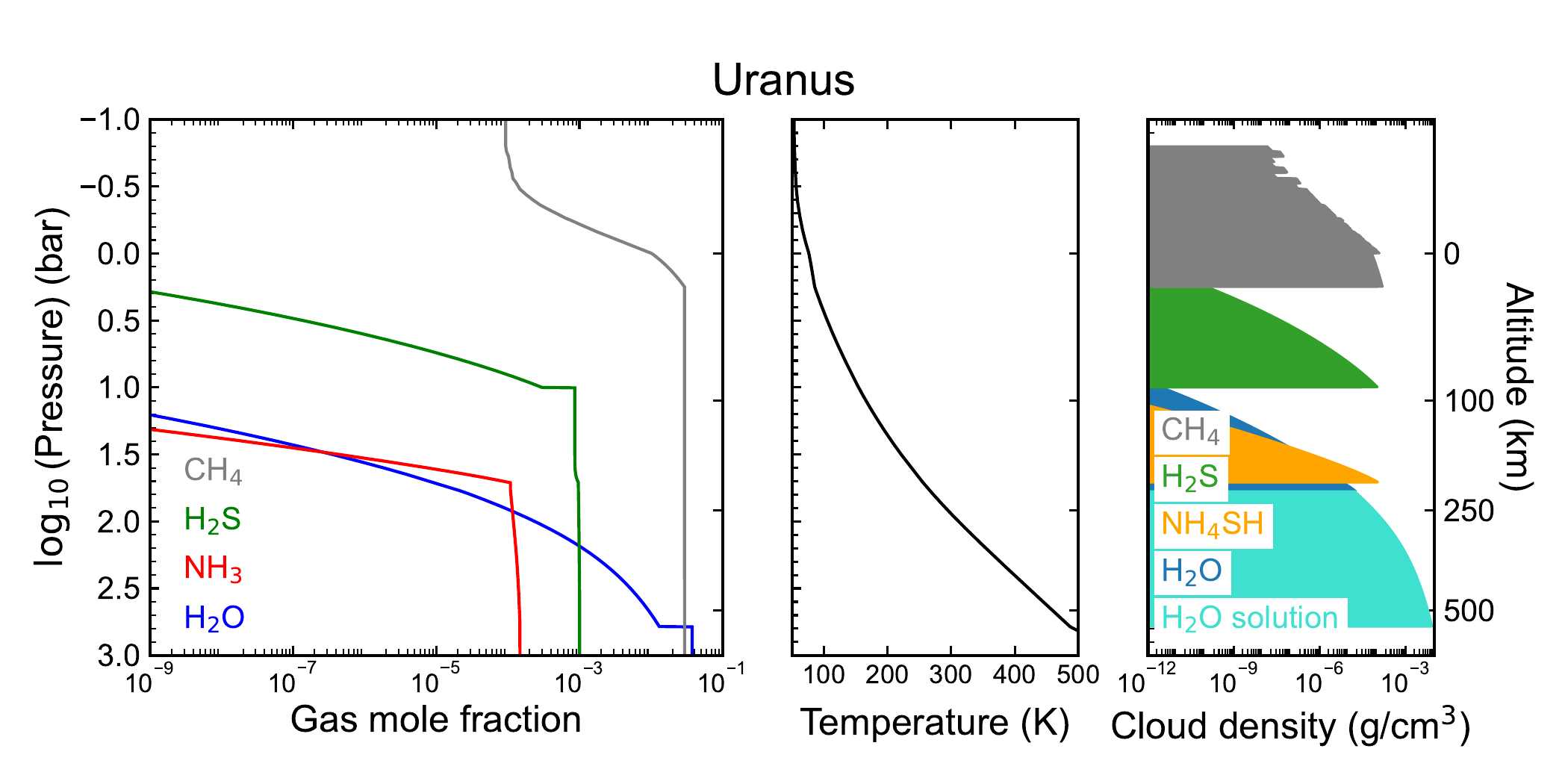} 
   \caption{Equilibrium cloud condensation models of the atmospheres of Saturn and Uranus. The input deep atmosphere compositions and relative humidities are given in the article text. Fluctuations in the upper clouds result from fluctuations in the radio occultation temperature profiles.}
   \label{fig:sat_ura}
\end{figure}

These relationships, while condensed, provide sufficient foundation for conducting first-order calculations relevant to mission formulation, which we will further elaborate in the following section. For expedience, we provide a relatively simple code\footnote{The ECCM code can be found at \url{https://github.com/abakins/ECCM_MRTM} \citep{alexakinsECCM_MRTMRelease2026}} for generating ECCM trace gas and temperature profiles as well as for computing forward model microwave brightness temperatures. Figure \ref{fig:sat_ura} shows the outputs of this code for nominal Saturn and Uranus models. For these atmospheres, we choose relative enrichments which are consistent with those described in the review of \cite{DePater2023} and use the same solar reference values (C/H$_2 = 5.90 \times 10^{-4}$, N/H$_2 = 1.48 \times 10^{-4}$, O/H$_2 = 1.07 \times 10^{-3}$, S/H$_2 = 2.89 \times 10^{-5}$, and P/H$_2 = 5.64 \times 10^{-7}$). For Saturn, deep abundances of H$_2$O, H$_2$S, and CH$_4$ are set to 10$\times$ the solar reference values, and NH$_3$ is set to 3$\times$. The relative humidity of NH$_3$ and H$_2$O are set at 70\% based on retrievals of NH$_3$ humidity from Cassini radiometer data by \cite{Laraia2013}. For Uranus,  H$_2$O and H$_2$S are set to 35$\times$ the solar reference values, CH$_4$ is set to 50$\times$, and NH$_3$ is set to 1$\times$. The relative humidity of H$_2$S and H$_2$O are set at 35\% based on retrievals of H$_2$S humidity from VLA observations by \cite{Akins2023}. Both of these models are based on information derived from prior studies at lower latitudes, with H$_2$O abundances being the most uncertain.

 \section{MWR Design Considerations} \label{sec:calcs}
 
With the fundamentals established, we can now discuss how different MWR instrument design choices will affect the capability of the measurement. For radiative transfer calculations, we use the model atmospheres for Saturn and Uranus (as a representative ice giant) in Figure \ref{fig:sat_ura} as a starting point. 

\subsection{Comparison with Earth-based Observations}
Given the resource limitations facing deep space missions, one may ask whether Earth-based microwave observatories are sufficiently powerful to achieve mission-scale science objectives. 
Earth-based alternatives to an orbital MWR include single-dish radio telescopes, such as the 100-m class Green Bank and Effelsberg observatories, or interferometer arrays, such as the Very Large Array (VLA). Potential future radio observatories include the Square Kilometer Array (SKA) and the next-generation VLA (ngVLA), whose proposed capabilities we will also consider. There are three factors which may impact the answer to this question: sensitivity and resolution, absolute accuracy, and viewing geometry/revisit time. 

The interferometric sensitivity equation, as discussed in detail by \cite{Wrobel1999, LeVine1990} is a more general formulation of the real aperture sensitivity equation (Equation \ref{eq:rase}) which accounts for the instantaneous field of view $\Omega_f$ and target spatial resolution $\Omega_s$ solid angles. This expression is based on the array collecting area and is as such a heuristic (since the image sensitivity will depend on the configuration of the array elements, approximately $\sqrt{N}$ for $N$ measured correlations), but it is usefully accurate when applied to most modern array telescopes.

 \begin{equation} \label{eq:sase}
\Delta T_A = \frac{\Omega_f}{\Omega_s}\frac{T_A + T_E}{\sqrt{NB \tau}}
\end{equation}

\begin{figure}[]
   \centering
   \includegraphics[width=0.75\linewidth]{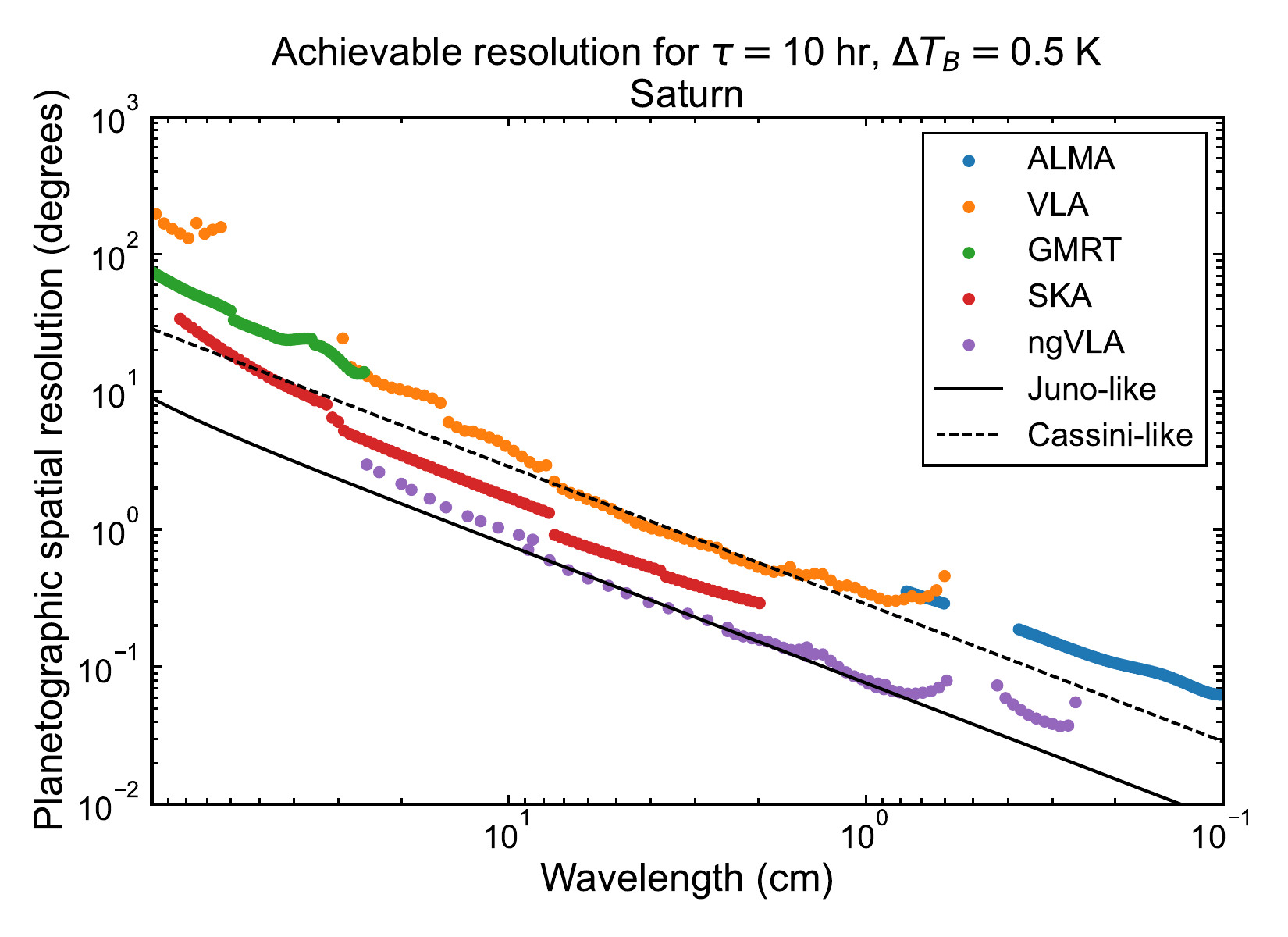}
      \includegraphics[width=0.75\linewidth]{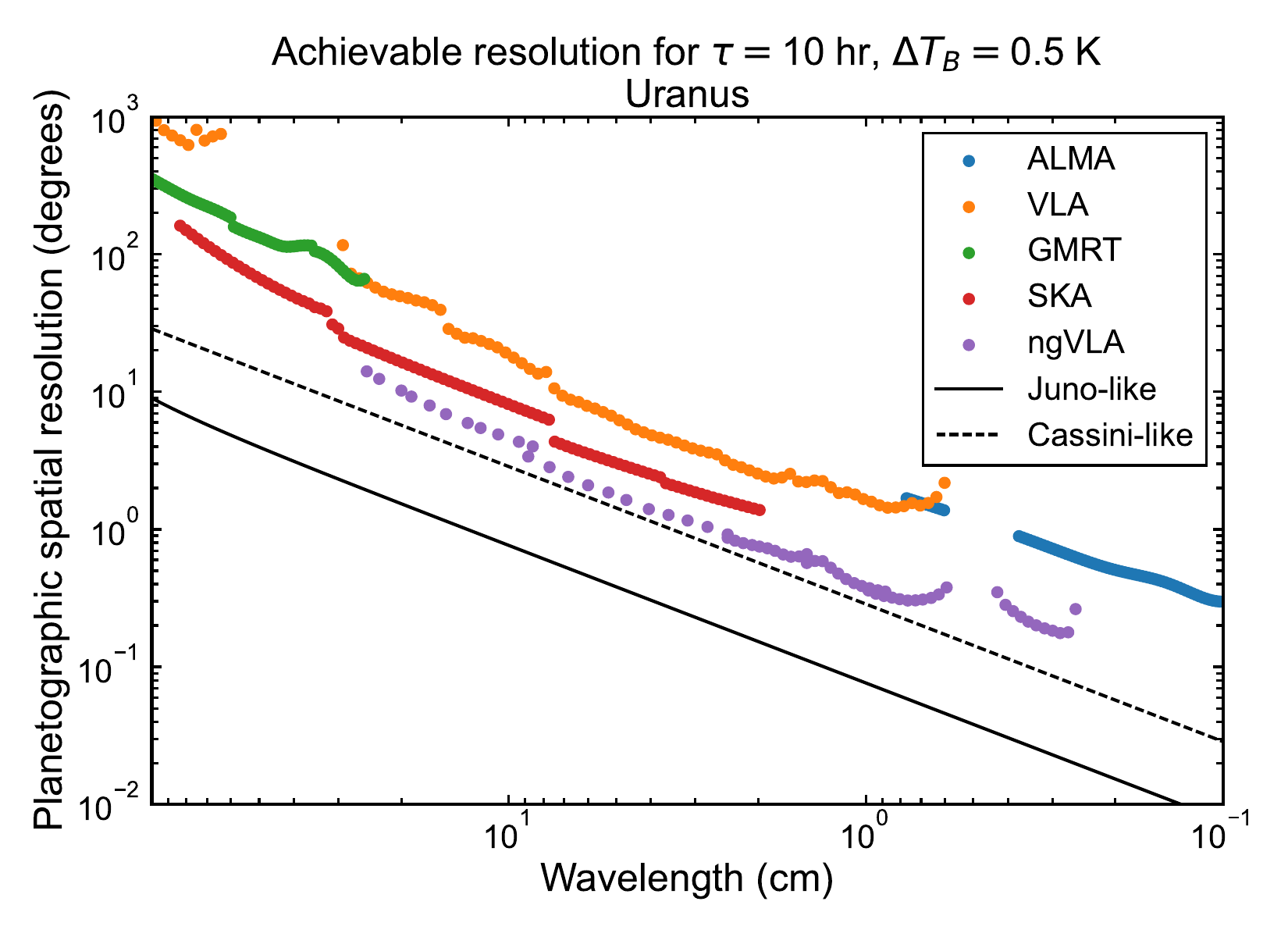}
   \caption{Achievable planetographic spatial (i.e. latitude/longitude) resolution with 0.5 K sensitivity for 10 hours on-source for several current (ALMA, VLA, GMRT) and planned (SKA, ngVLA) radio telescope arrays for Saturn and Uranus. The black curves show the spatial resolution obtainable by MWR's in Juno-like and Cassini-like orbits with comparable antenna sizes (Juno channel 2, Cassini HGA). Spatial resolutions are computed using small-angle approximations; values in excess of 120 degrees are non-physical and should be considered to resolve only the disk-integrated emission.}
   \label{fig:gbmwr}
\end{figure}

From this relationship, we can see that the sensitivity and spatial resolution of interferometric observations for equally sized arrays are inversely proportional. Improvements in spatial resolution at comparable sensitivity requiring an increase in the number of array correlation measurements, observing bandwidth, and/or time on source. Figure \ref{fig:gbmwr} uses Equation \ref{eq:sase} to connect the achievable sensitivity and resolution of Earth-based telescopes with orbiting MWR's. To bracket the calculation, we assume a target sensitivity of 0.5 K (approximate per-integration sensitivity for contemporary MWR's) and ten hours of observing time. The observing bandwidths as a function of wavelength are set at the maximum per-band continuum bandwidth used for multi-frequency synthesis imaging (not the apparent wavelength spacing of the scatter points). Array system temperatures are taken from available measurements and array design documents \citep{Braun2019,Selina2017,Remijan2019}. The sold and dashed lines show achievable resolution for Juno-like (Channel 2-equivalent 75 cm aperture with orbit periapse of $1.1R_p$, where $R_p$ is the 1-bar radius of the planet) and Cassini-like (HGA-equivalent 4 m aperture, orbit periapse of $3R_p$) MWRs. For Saturn, current observatories can achieve comparable sensitivity/resolution to a Cassini-like radiometer, whereas future observatories, if successfully completed, would achieve performance comparable to Juno. This statement applies mostly to intermediate wavelengths, as the long-wavelength performance of potential next-generation observatories still falls short of orbital MWR capabilities. For Uranus, the difference is more stark; future observatories will not be able achieve comparable performance to orbital instruments. From a sensitivity perspective, long-wavelength observations should be prioritized in the formulation of future MWRs, as prior meter-wavelength observations of Saturn and recent results for Uranus have achieved only disk-integrated temperatures with poor precision (Appendix \ref{app:A}).  

Of similar importance to precision is the matter of absolute brightness temperature calibration accuracy, which is determined via cold and hot reference points. The cold reference for Earth-based observatories and space-based MWRs is extrasolar background emission, with the cosmic component being isotropic to sub-milliKelvin level and the low-frequency galactic continuum/HI emission having been mapped to good accuracy \citep{anderson_radio_2025}. For Earth-based observatories, the hot references are near-proximal observations of astrophysical reference standards. Achievable accuracies with this approach are on the order of 3-10\% of the emission temperature \citep{Perley2017, Farren2021}, with decreasing accuracy at the longest and shortest wavelengths. The absolute flux densities of extrasolar calibrators are themselves determined based on calibration against models of Mars ($>$2 GHz), strong sources such as Cygnus A ($<$ 2 GHz), and observations from space-based observatories such as Planck and WMAP. At millimeter wavelengths, solar system objects are often used as reference calibrators when high absolute accuracy is necessary \citep{Butler2012}. The main obstacles to the absolute calibration of ground-based observatories are Earth's troposphere and ionosphere. For frequencies higher than 1 GHz, absolute calibration accuracy generally drives measurement uncertainty. Spacecraft MWR's use fabricated calibration sources as the hot reference, including warm external targets (for scanning radiometer systems) and internal reference loads and noise diodes. While internal noise-injection and external water vapor radiometer calibration approaches are employed at some observatories, direct amplitude calibrations are still poorer than those inferred from sky calibrator observations due to atmospheric variability. As such, space-based MWR measurements are capable of obtaining higher brightness temperature accuracy than ground-based observations (order 1-2\%). Importantly, orbital MWR's, particularly those with Juno's spinning design, can also measure emission at different look angles with order 0.1\% precision in limb-darkening; these measurements are relative and therefore are more robust against absolute calibration offsets. 

Finally, there is the matter of viewing geometry and revisit time. The main advantage of Earth-based observatories as compared to orbital MWR systems is their ability to contemporaneously image an entire hemisphere (or even the full disk for imaging through an entire rotation, \cite{Sault2004, DePater2016}) of a planet. In most other aspects, space-based MWR's are superior. Earth-based observations are inherently limited in the range of observable latitudes for low obliquity objects, although the polar regions of giant planets can be observed at certain seasons.  Earth-based observations are also incapable of making near-co-located observations at multiple emission angles. For spacecraft observations, the observation revisit time is determined by the spacecraft orbit and pointing characteristics, whereas for ground-based observations, revisit time is determined by the viewable window from Earth and from observatory scheduling pressure.

\subsection{Wavelength and Angular Spectrum Coverage}
\begin{figure}[htbp]
   \centering
   \includegraphics[width=0.75\linewidth]{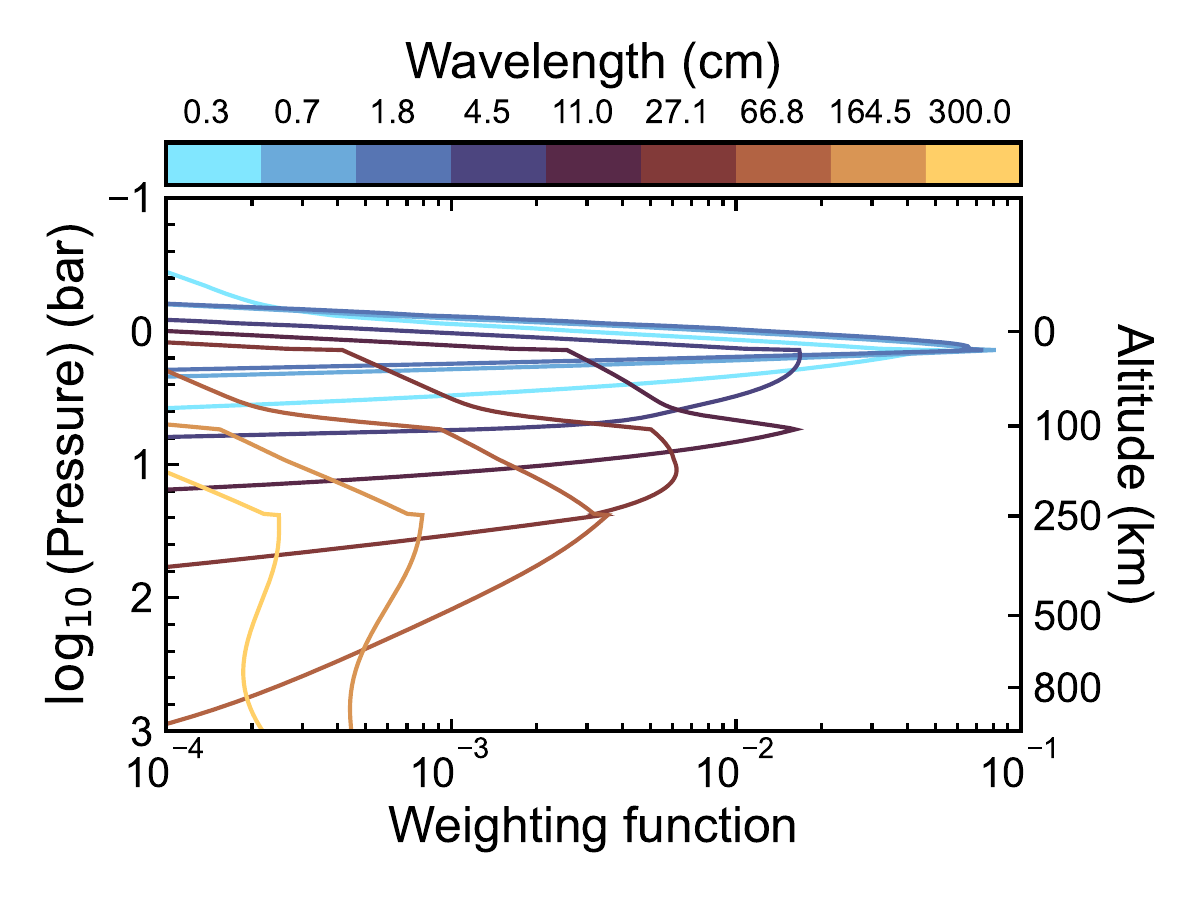} 
    \includegraphics[width=0.75\linewidth]{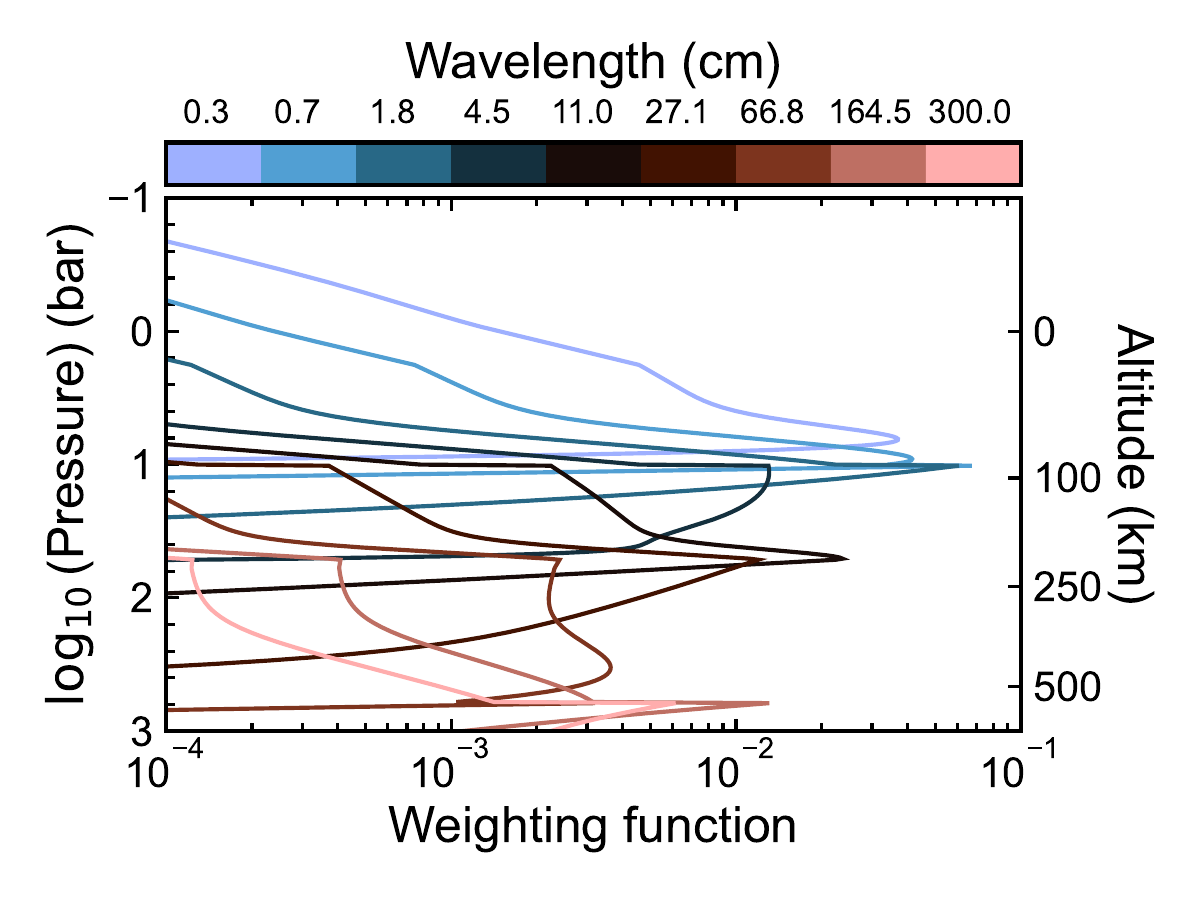}
   \caption{Example weighting functions for selected microwave frequencies (logarithmically spaced between 3 mm - 3 m) corresponding to the model atmospheres for Saturn (top) and Uranus (bottom) from Figure \ref{fig:sat_ura}.}
   \label{fig:weight}
\end{figure}

The composition and temperature of giant planet atmospheres affect the wavelength and angular spectrum of their thermal emission. We might therefore ask about the minimal set of frequencies and angles which should be observed. While the answer to this question depends on the science goal, we present here some calculations which qualitatively illustrate the impact of different choices. Juno MWR, for example, made observations at six wavelengths (50, 24, 12, 6, 3, 1.4 cm) and leveraged the spin-stabilized spacecraft to observe at many emission angles \citep{Janssen2017}. In absence of any knowledge about the atmosphere being observed, the octave-interval approach adopted for the center frequencies of Juno MWR is likely a reasonable one for approximately homogeneous vertical resolution. This results from the general tendency of the absorption coefficient for an arbitrary medium to increase with the square of frequency and the fact that the weighting function term in the radiative transfer equation is the derivative of atmospheric transmissivity with height \citep{Ulaby}. An alternative approach is to make channel selections informed by nominal atmospheric models and radiative transfer calculations. This approach is relevant when the vertical distribution of atmospheric opacity deviates from a smooth continuum. This is in fact the case for the giant planets; ECCM's predict substantial changes in atmospheric composition at the cloud boundaries, and they result in significant transitions in opacity and highly peaked weighting functions, as shown in Figure \ref{fig:weight}.   

\begin{figure}[htbp]
   \centering
   \includegraphics[width=\linewidth]{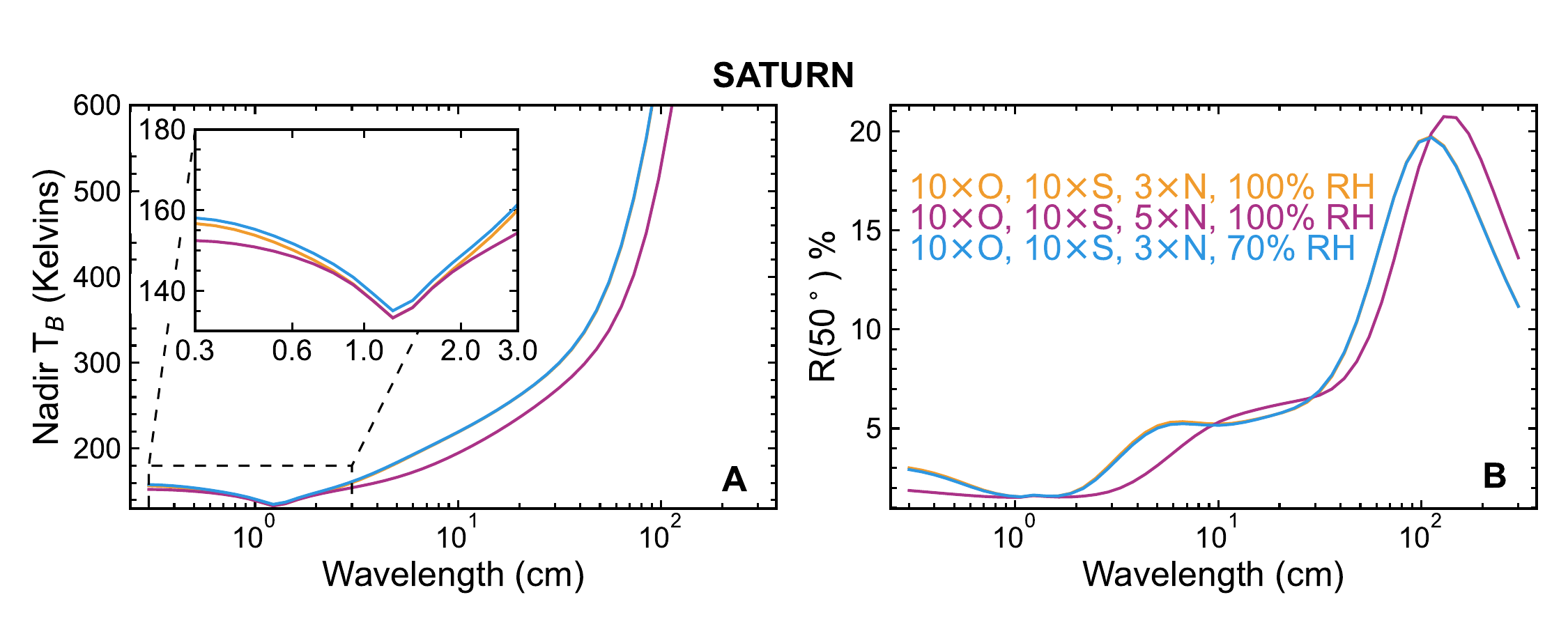} 
    \includegraphics[width=\linewidth]{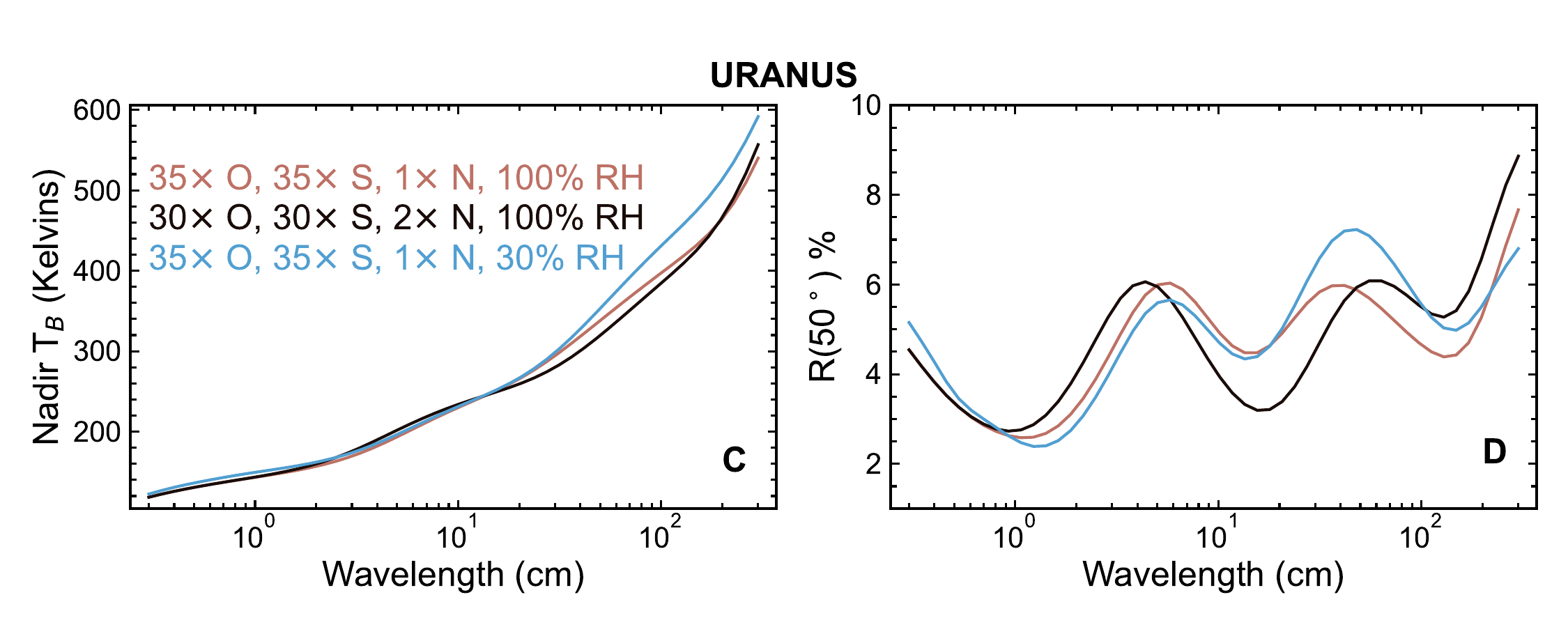}
   \caption{Nadir brightness temperature (A, C) and limb-darkening spectra (B, D, $R(\theta)=|T_B(0)-T_B(50^\circ)|/T_B(0)$) for atmospheric models of Saturn and Uranus in Figure \ref{fig:sat_ura}. The effects of some arbitrary perturbations to this model are also illustrated (see figure text). }
\label{fig:sat_ura_refspec}
\end{figure}

We assess the sensitivity of the microwave spectrum to the abundances of atmospheric trace gases and their relative humidities in the condensational cloud by perturbing the input parameters to the model atmospheres in Figure \ref{fig:sat_ura} and generating spectra; this allows some insight into which channels we can use to best estimate the deep abundances of N, S, and O, and the precision to which the measurements need to be made. \cite{Molter2021} and \cite{Tollefson2021} provide similar curves in their respective analyses of Uranus and Neptune VLA and ALMA observations. We show in Figure \ref{fig:sat_ura_refspec} nadir brightness temperature and limb-darkening ($R(\theta)=|T_B(0)-T_B(\theta)|/T_B(0)$) spectra for the atmosphere models in Figure \ref{fig:sat_ura}, as well as some illustrative adjustments. Subsequent figures can be interpreted as deviations from the reference model (green curves). 

\begin{figure}[htbp]
   \centering
   \includegraphics[width=0.8\linewidth]{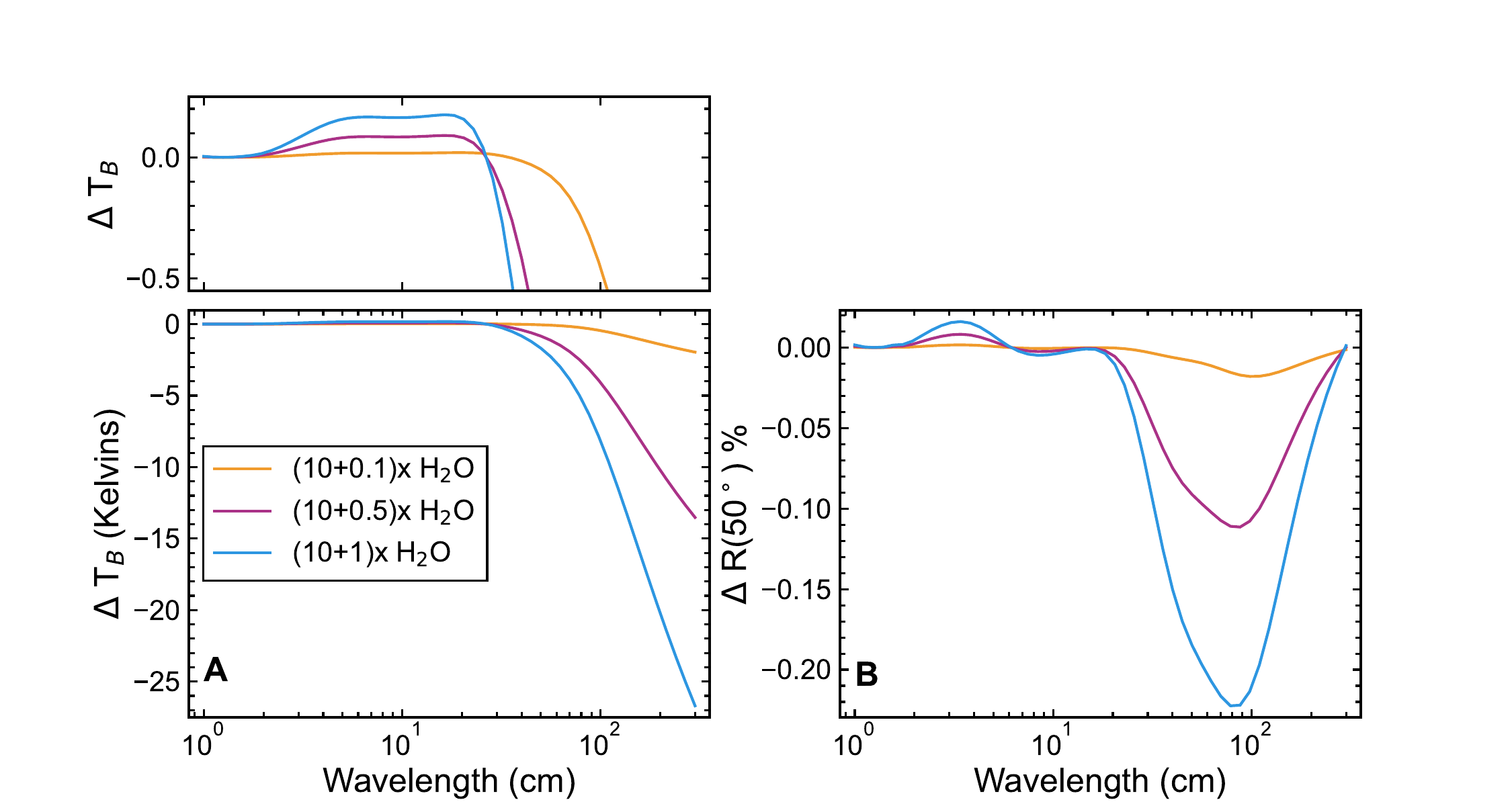} 
    \includegraphics[width=0.8\linewidth]{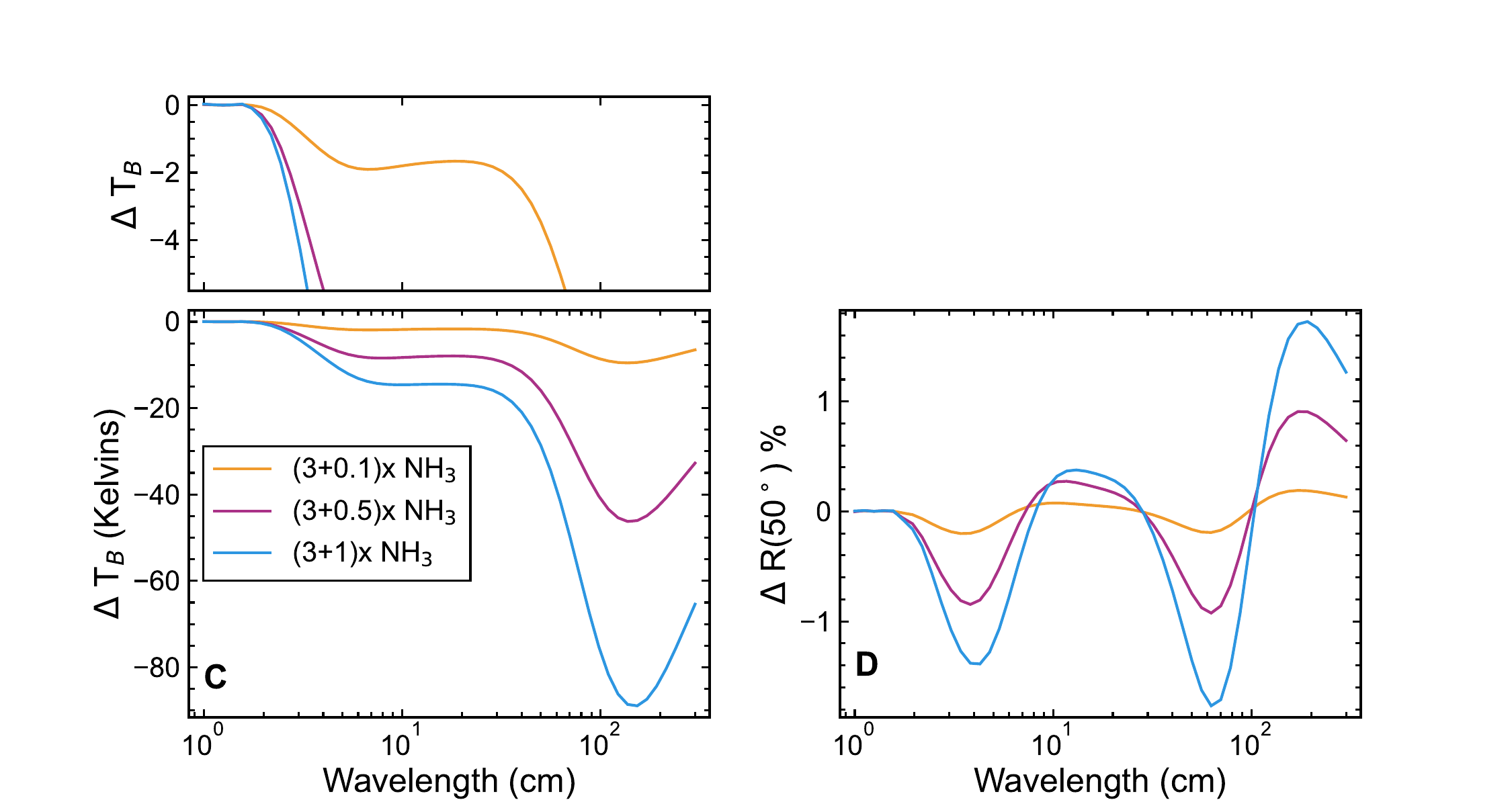}
   \caption{Effect of perturbing the deep abundance of H$_2$O (A, B) and NH$_3$ (C, D) on Saturn's microwave nadir spectrum (left) and 50-degree limb-darkening (right). Subplots above the $\Delta T_B$ spectra show the same curves with different axis limits.}
\label{fig:sat_perturb_singles}
\end{figure}

\begin{figure}[htbp]
   \centering
   \includegraphics[width=0.8\linewidth]{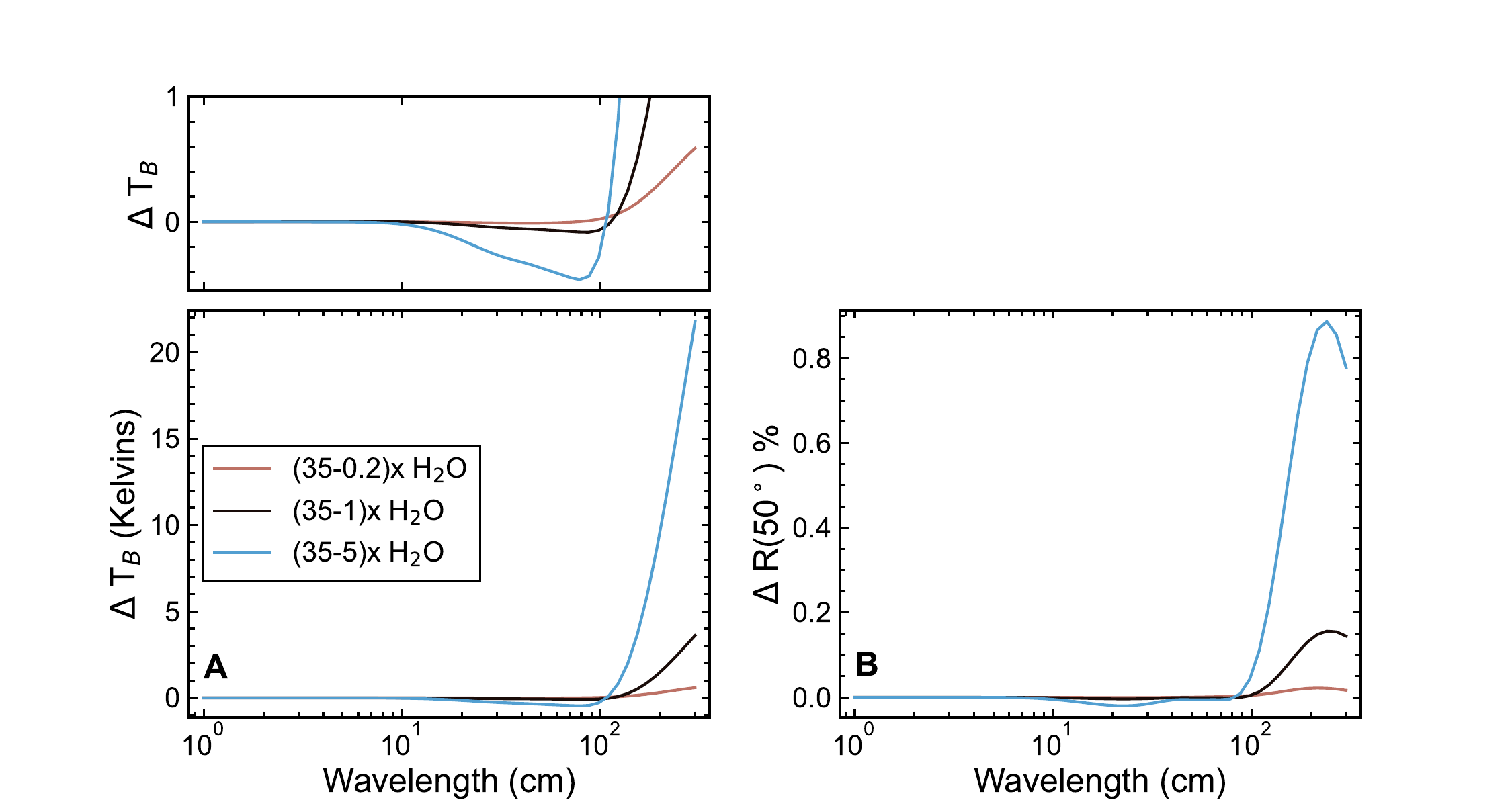} 
    \includegraphics[width=0.8\linewidth]{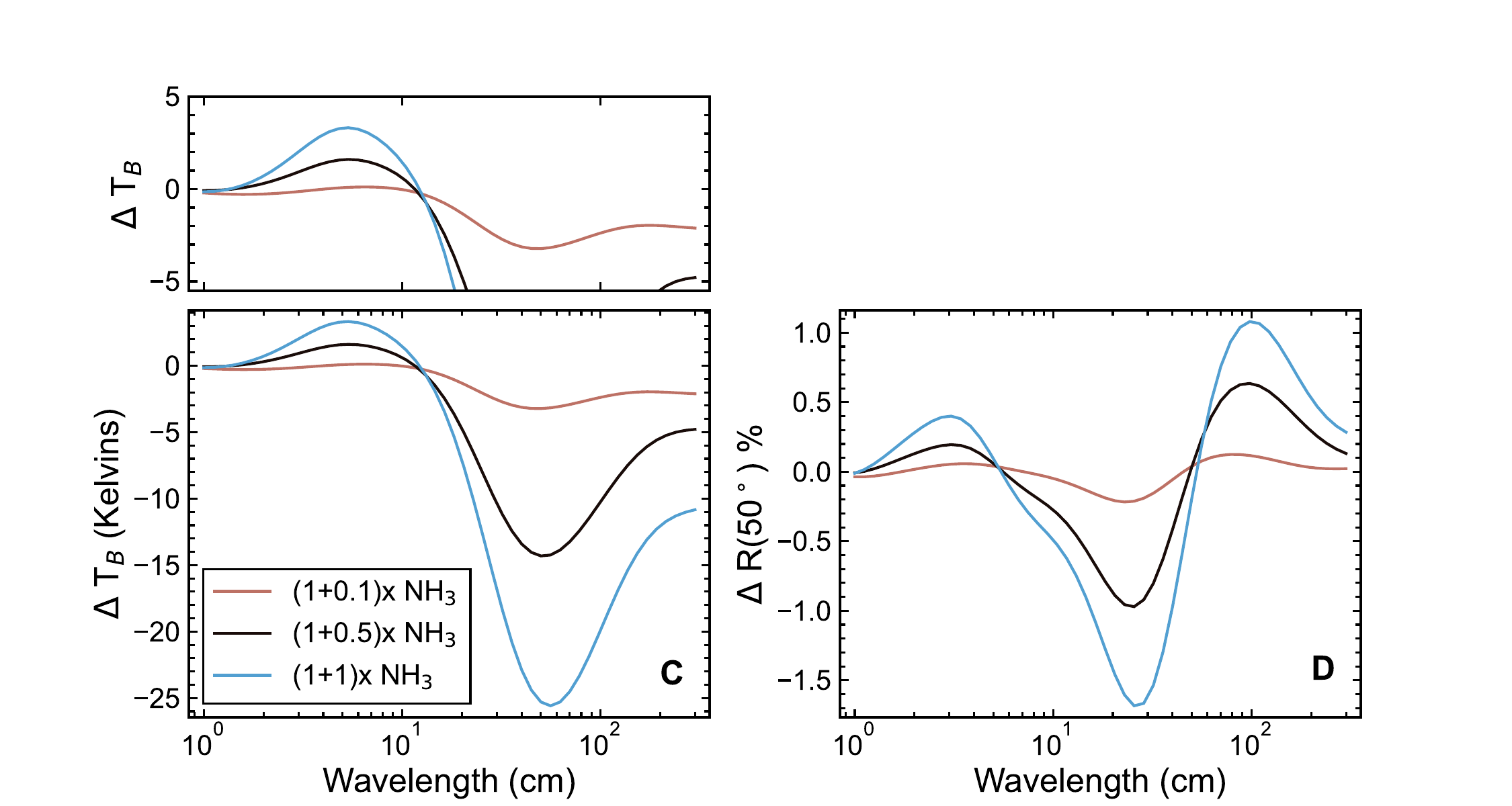}
    \includegraphics[width=0.8\linewidth]{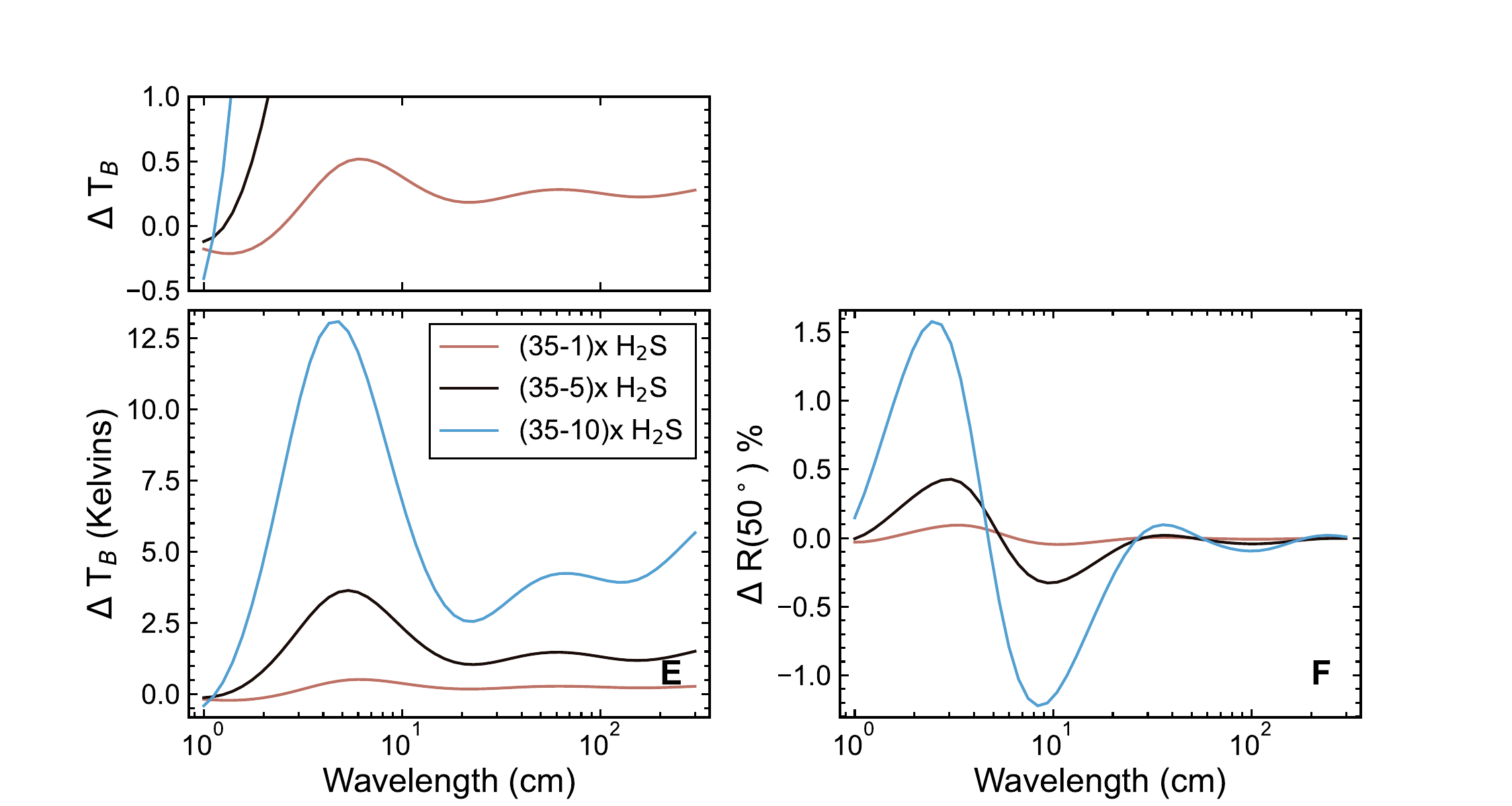}
   \caption{Effect of perturbing the deep abundance of H$_2$O (top), NH$_3$ (middle), and H$_2$S (bottom) on Uranus' microwave nadir spectrum (left) and 50-degree limb-darkening (right).  }
\label{fig:ura_perturb_singles}
\end{figure}

\begin{figure}[htbp]
   \centering
   \includegraphics[width=\linewidth]{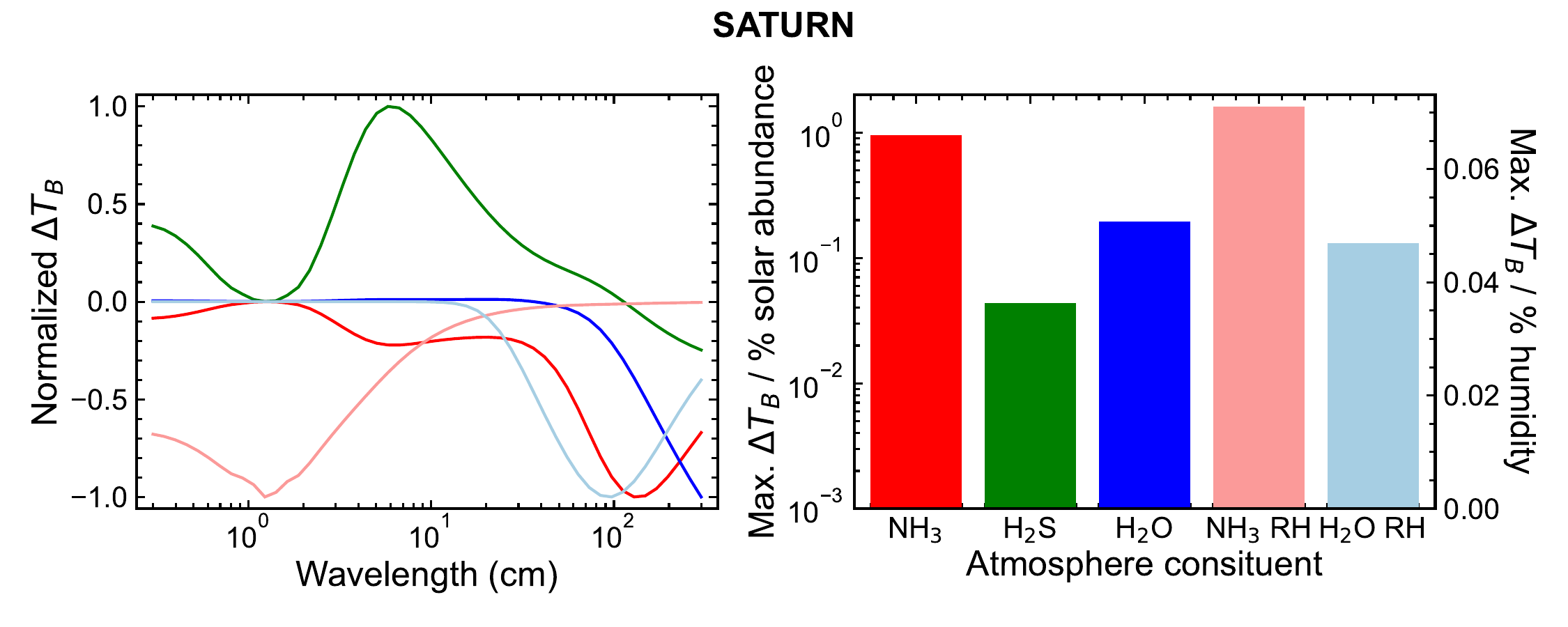} 
    \includegraphics[width=\linewidth]{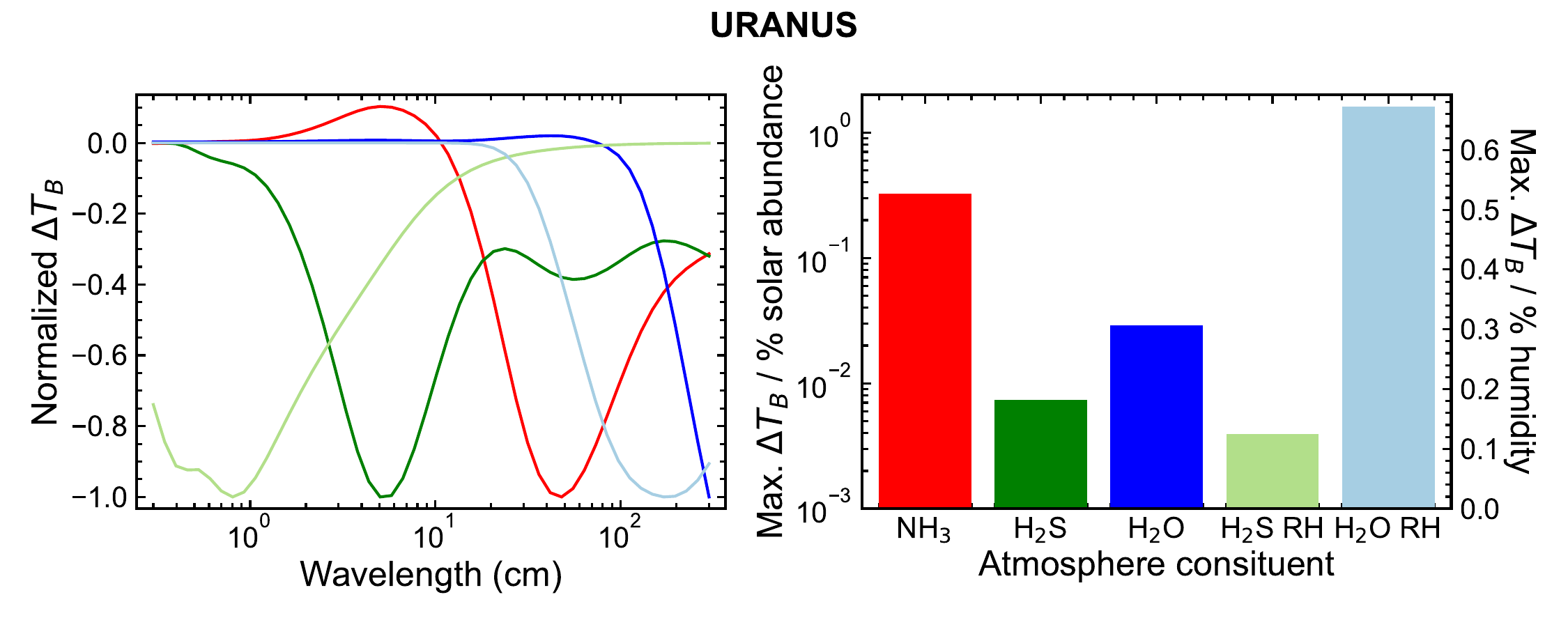}
   \caption{Normalized change in brightness temperature as a function of wavelength (left) and peak values for the un-normalized curves (right) resulting from local linear perturbations of atmospheric composition for Saturn (top) and Uranus (bottom). The peak value is reported as a change in brightness temperature per \% change in solar-relative abundance or per \% change in relative humidity.}
\label{fig:specpert}
\end{figure}

Figures \ref{fig:sat_perturb_singles} and \ref{fig:ura_perturb_singles} show the effects of discrete perturbations in composition on the spectrum. We note that the effect of changing the deep abundances of H$_2$S or NH$_3$ will affect the other through the condensation of the NH$_4$SH cloud. The same applies to the H$_2$O-NH$_3$-H$_2$S solution cloud, although the effect is less significant (see Figure \ref{fig:sat_ura}). Figure \ref{fig:specpert} shows the normalized spectra and peak values for reference spectra in which quantities of interest are perturbed by 1\% .This figure can be interpreted as defining the radiometric sensitivity necessary for an orbital MWR to resolve a 1\% change in solar-relative abundance or relative humidity. Due to its considerable microwave opacity, small changes in NH$_3$ can be resolved at high precision, order 10\% solar. Inferences of H$_2$S and H$_2$O abundances at a comparable precision would require an order of magnitude increase in precision or averaging over many orbits.

With regards to composition, it is important to keep in mind the effect of atmospheric thermal structure as a confounding factor, particularly in the case of weaker absorbers like H$_2$S gas. \cite{Janssen2005} and \cite{DePater2005} discussed specifically the challenge of resolving between different O/H compositional scenarios for Jupiter from an MWR instrument. \cite{DePater2005} specifically hypothesized that it would not be possible ``to distinguish an atmosphere with solar water from one with $\sim$5$\times$ solar, and neither does it seem possible to distinguish 10$\times$ from $\sim$20$\times$ solar,'' whereas \cite{Janssen2005} posited that the O/H ratio could be resolved to be either $< $1, $\sim$3, or $>$9. Two different constraints were obtained from actual Juno measurements (both in the equatorial zone): 2.7$^{+2.4}_{-1.7}$ by \cite{Li2020} based on varying H$_2$O directly, and 4.9$^{+3.4}_{-3.4}$ by \cite{Li2024} based on the best-fit atmospheric temperature gradient. These retrievals, combined with the context of CO thermochemical modeling \citep{hyder_supersolar_2025}, suggest that Janssen's intermediate outcome ($\sim 3 \times$ solar) is the true value. In the case of Jupiter's H$_2$S, the results of \cite{Moeckel2023} suggest that its abundance is perhaps less than that inferred by Galileo measurements, but its abundance in general is too low to strongly constrain. 

What do these results mean for Saturn and Uranus? In both cases, the atmosphere is colder at the same pressure levels, meaning that cloud condensation altitudes are deeper than on Jupiter. This is mostly a concern for inferring O/H, and is of greatest consequence for Uranus. For the ice giants, long-wavelength MWR measurements may only be able to rule out a low O/H enrichment, whereas a ``low-intermediate-high'' hypothesis system (like that posited by \cite{Janssen2005}) can likely still apply to Saturn if a sufficiently long-wavelength channel is employed. In either case, longer wavelengths than 50 cm should be employed for observations of Saturn and Uranus if O/H is to be inferred. If aperture size becomes prohibitive to achieve high spatial resolution at meter-wavelengths, even disk-integrated measurements using an electrically small antenna (i.e. with diameter $<< \lambda$) at e.g. 3 m would be of great utility. For H$_2$S, the increased condensation depth is offset by the likely greater atmospheric abundances, particularly for the ice giants. MWR observations beyond Jupiter are much more likely to usefully retrieve S/H. 

\begin{figure*}[htbp]
   \centering
   \includegraphics[width=0.45\linewidth]{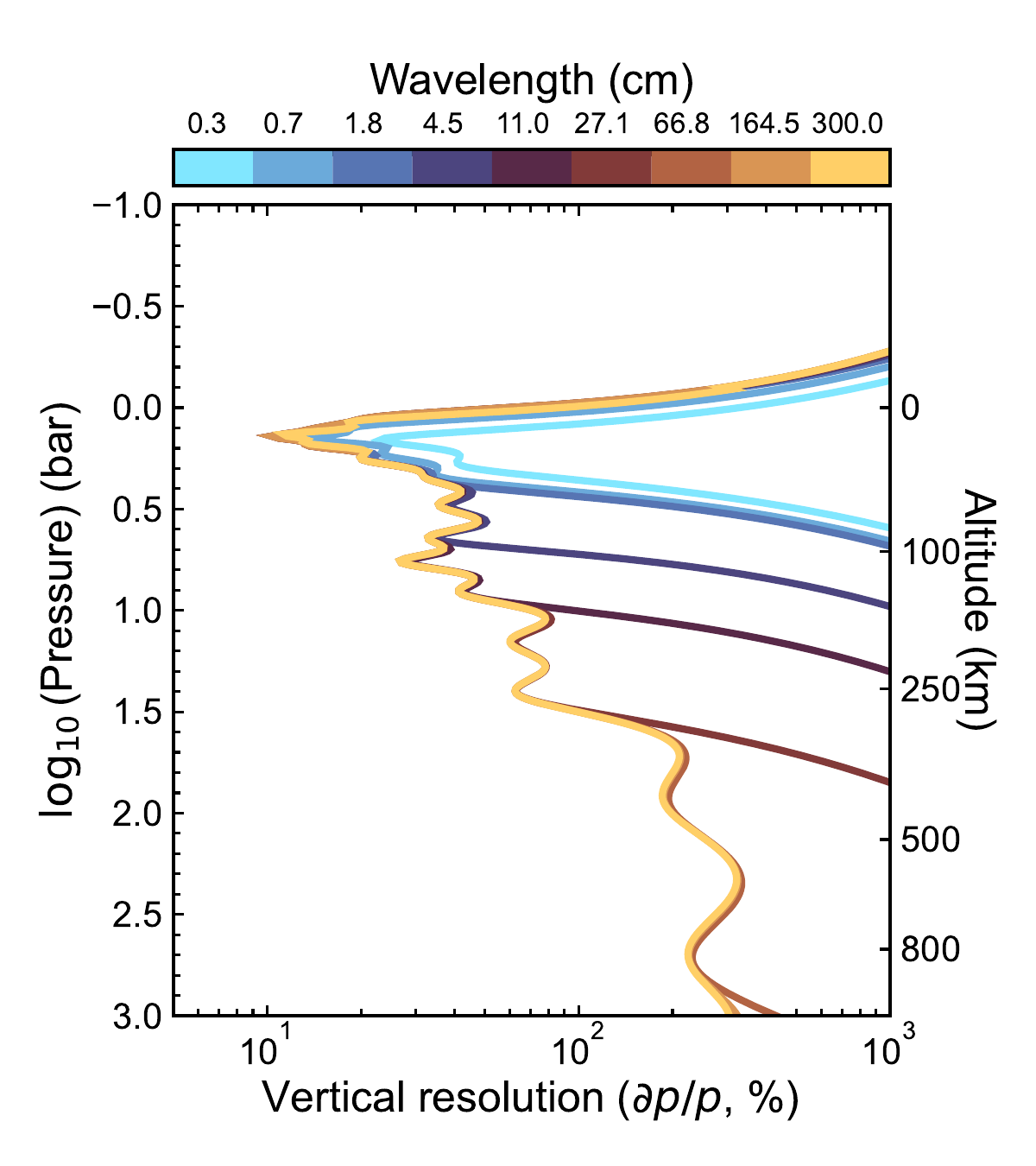} 
    \includegraphics[width=0.45\linewidth]{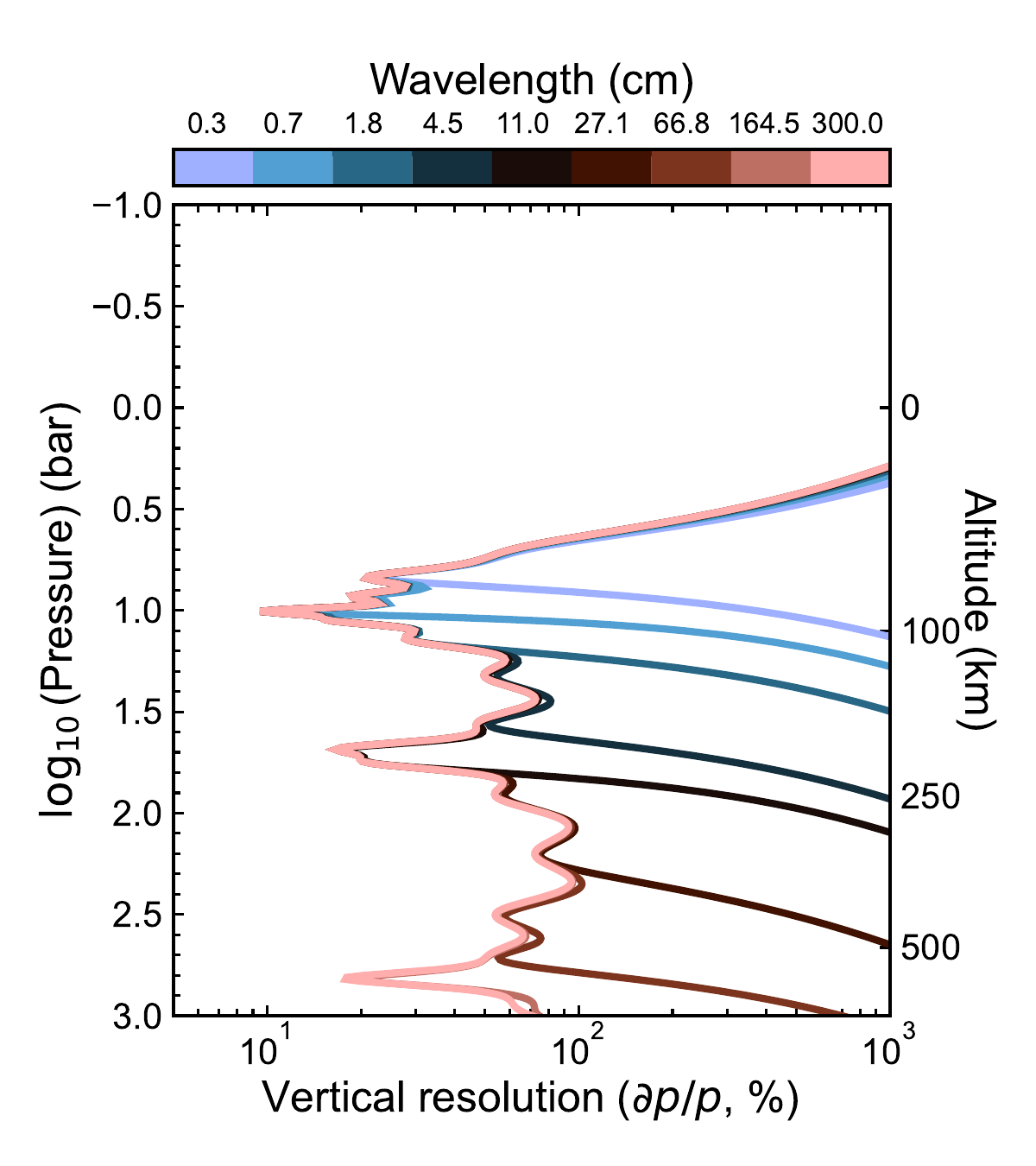} 
   \caption{Vertical resolution (Backus-Gilbert spread) for MWR measurements of the atmospheres of Saturn and Uranus (Figure \ref{fig:sat_ura})  using channel combinations from Figure \ref{fig:weight}, starting from the shortest wavelength and progressively adding channels. Nadir and 50\% incidence angles are included in the spread metric calculation.}
   \label{fig:vert_res}
\end{figure*}

While less relevant for inferring global deep abundances, the vertical resolution achieved by a particular combination of MWR channel frequencies and look angles is important for the assessment of atmospheric dynamics. We illustrate in Figure \ref{fig:vert_res} calculations of the Backus-Gilbert spread, which is a metric for vertical resolution, for different combinations of channels in Figure \ref{fig:weight}. For a collection of temperature weighting functions $\mathbf{K}$, the spread metric $V(r)$ is computed as

\begin{align} 
\begin{split}
&V(r)=\frac{1}{\mathbf{k}^T\mathbf{Q}^{-1}(r)\mathbf{k}} \\ 
&k_i=\int{K_i\left(r'\right)dr'} \\ 
&Q_{ij}\left(r\right)=12\int{(r-r')}^2K_i\left(r'\right)K_j\left(r'\right)dr
\end{split}
\end{align} 

We note that the calculation of the spread metric as described here minimizes vertical resolution, whereas a trade-off parameter can be included to balance the resulting error; the interested reader is referred to \cite{Rodgers2000}, Chapter 4, for a more complete discussion.

\cite{Rodgers2000} also defines metrics for the information content of measurements made with different observing systems. We calculate here the degrees of freedom for signal ($d_s$, or the number of linearly independent quantities measured) and Shannon's information content metric ($H$, or the logarithm of the number of possible states that the measurement can discriminate between) for different frequency/angle combinations for observations of planets beyond Jupiter. These metrics can be computed from the singular values $\lambda$ for the singular value decomposition of the weighting function matrix $\mathbf{K}$. 

\begin{align} 
\begin{split}
& d_s=\sum\frac{{\lambda_i}^2}{{{1+\lambda}_i}^2} \\ 
& H =\frac{1}{2} \sum \log{(1 + \lambda_i^2)} 
 \end{split} 
 \end{align}
 
\begin{figure}[htbp]
   \centering
      \includegraphics[width=0.75\linewidth]{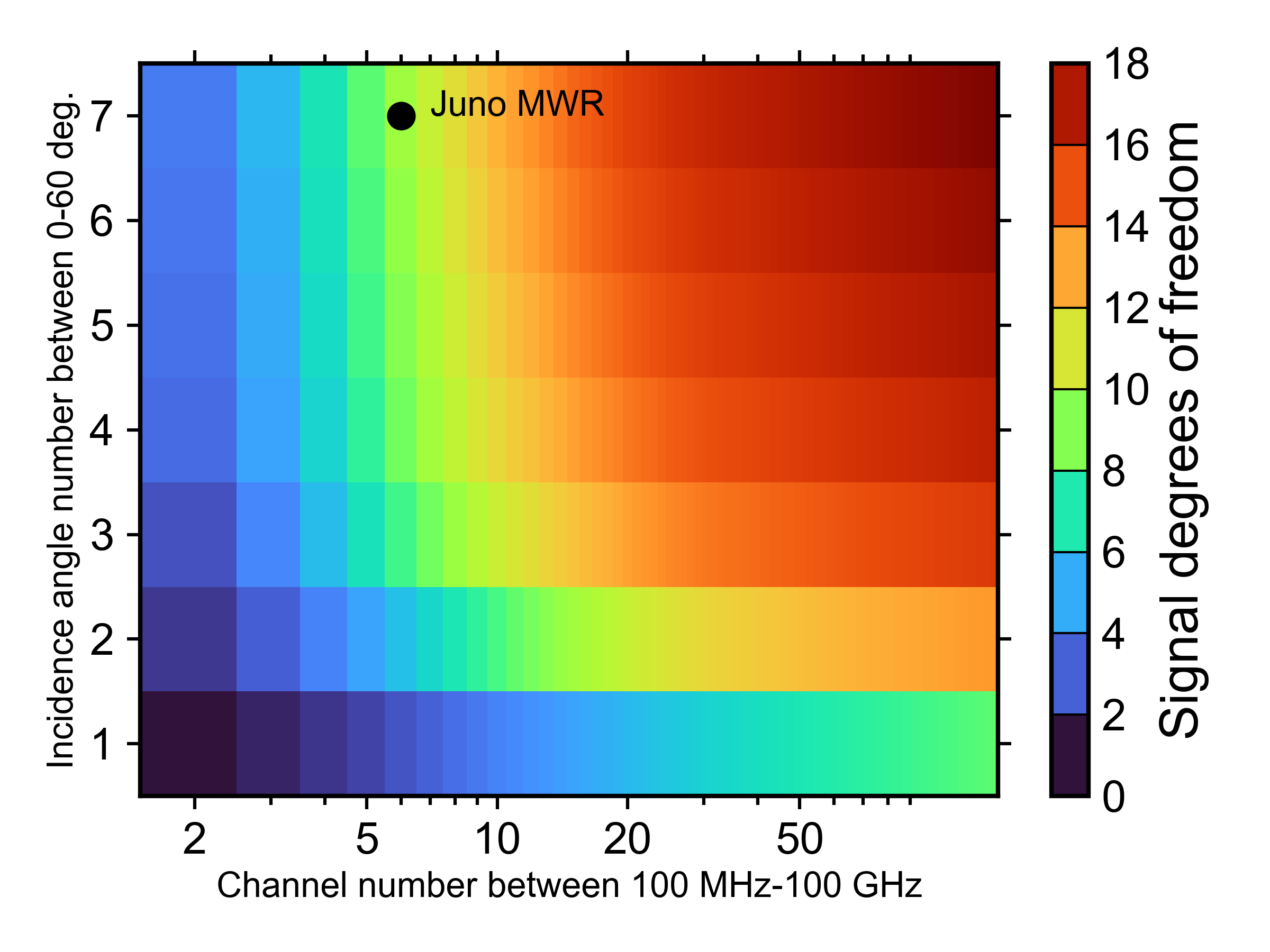} 
   \includegraphics[width=0.75\linewidth]{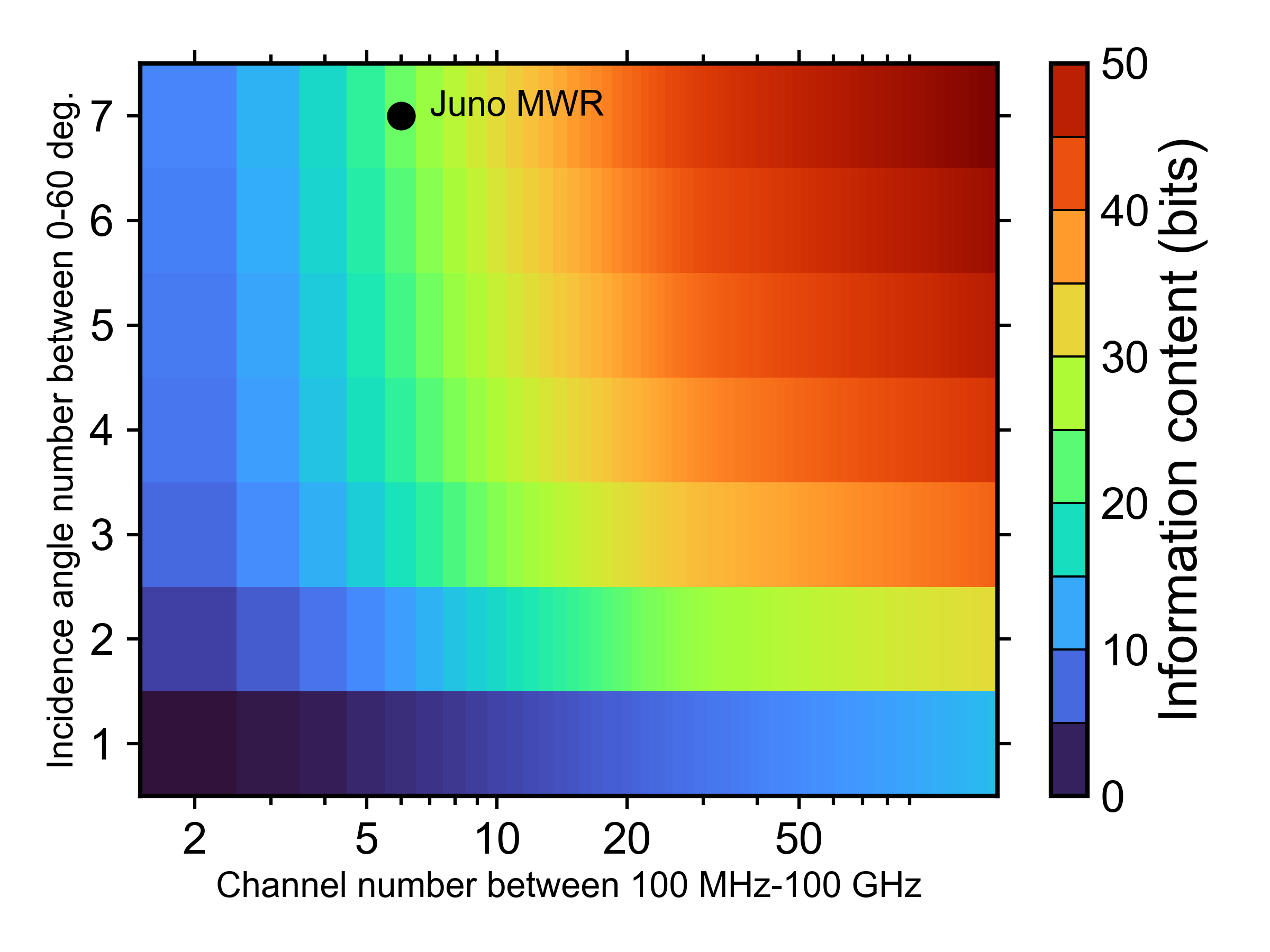} 
   \caption{Degrees of freedom for signal (top) and Shannon information content (bottom) for MWR sounding of Saturn's atmosphere using a variable number of channels between 0.3 and 300 cm and different incidence angles. The region of this parameter space corresponding to Juno MWR is identified explicitly. We note that Juno MWR observes $>$ 7 incidence angles, but for the relevant noise levels, the additional information provided by each further measured angle decreases past this point.}
\label{fig:info}
\end{figure}

The covariance matrix $\mathbf{S}_\epsilon$ terms are computed using Equation \ref{eq:rase} with $T_E = 1000$ K, and for the nadir observations, an additional 2\% error due absolute calibration uncertainty is included. Figure \ref{fig:info} shows the values of $d_s$ and $H$ computed for our nominal Saturn atmosphere in Figure \ref{fig:sat_ura}. 
 
Taken at face value, the implication of Figure \ref{fig:info} is that different combinations of spectral and angular coverage can recover similar amounts of information regarding the thermal state of the atmosphere. Juno MWR made measurements at 6 discrete frequencies over a wide range of angles, but perhaps a non-spinning spacecraft employing well-calibrated receivers with higher spectral resolution could obtain results which are similarly informative. A more complete analysis is required, however, before this can be confidently stated. For example, it may be possible that multi-angle observations are in fact more useful than multi-spectral observations due to the effect of longer pathlengths at larger incidence angles effectively narrowing the width of the weighting function. It is also worth mentioning here that, although prior flown MWR systems have employed multiple narrow-bandwidth channels with independent antenna and receiver systems, this is not strictly necessary. Technological developments since Juno was formulated have enabled the development of ultra-wideband microwave radiometer antennas and advanced digital receiver systems that have the potential to reduce instrument footprint, improve spectral coverage, and enable novel imaging architectures \citep{misraPlanetaryBoundaryLayer2024, Johnson2021, rodriguez-fernandez2024}.

While the presented information content metrics have formal definitions, our discussion here has a qualitative bent due to the ill-posed nature of the problem at hand. Precise and accurate retrievals of either atmospheric temperature or composition from microwave observations require that the other quantity is known with some confidence. This is also the case for millimeter and infrared spectral line observations, but the required confidence in the knowledge of the auxiliary quantity is higher for microwave observations due to the quasi-continuum nature of the spectrum (except e.g. in the vicinity of low-frequency NH\textsubscript3 and H\textsubscript{2}O lines). Prior to Juno and Cassini, ECCM models were assumed to provide sufficient information regarding the connection between temperature and composition so as to render this problem more tractable. We have since discovered that these models are less consistent than expected with the actual observations, from which we can conclude that Jupiter and Saturn's deep atmospheres are not strictly adiabatic and that substantial inhomogeneities exist in the distributions of absorbing trace gases. Future MWR experiments would benefit from the following efforts: (1) a more rigorous assessment of how different combinations of spectral and angular coverage in MWR measurements can be applied to disentangle the degeneracies of the retrieval problem, and (2) the development of models which are capable of more accurately accounting for the complex interdependencies of atmospheric dynamics and state while retaining some of the computational expedience that makes ECCMs attractive. 

 \subsection{How do we size MWR antenna apertures?}
 
\begin{figure}[htbp]
   \centering
	\includegraphics[width=0.75\linewidth]{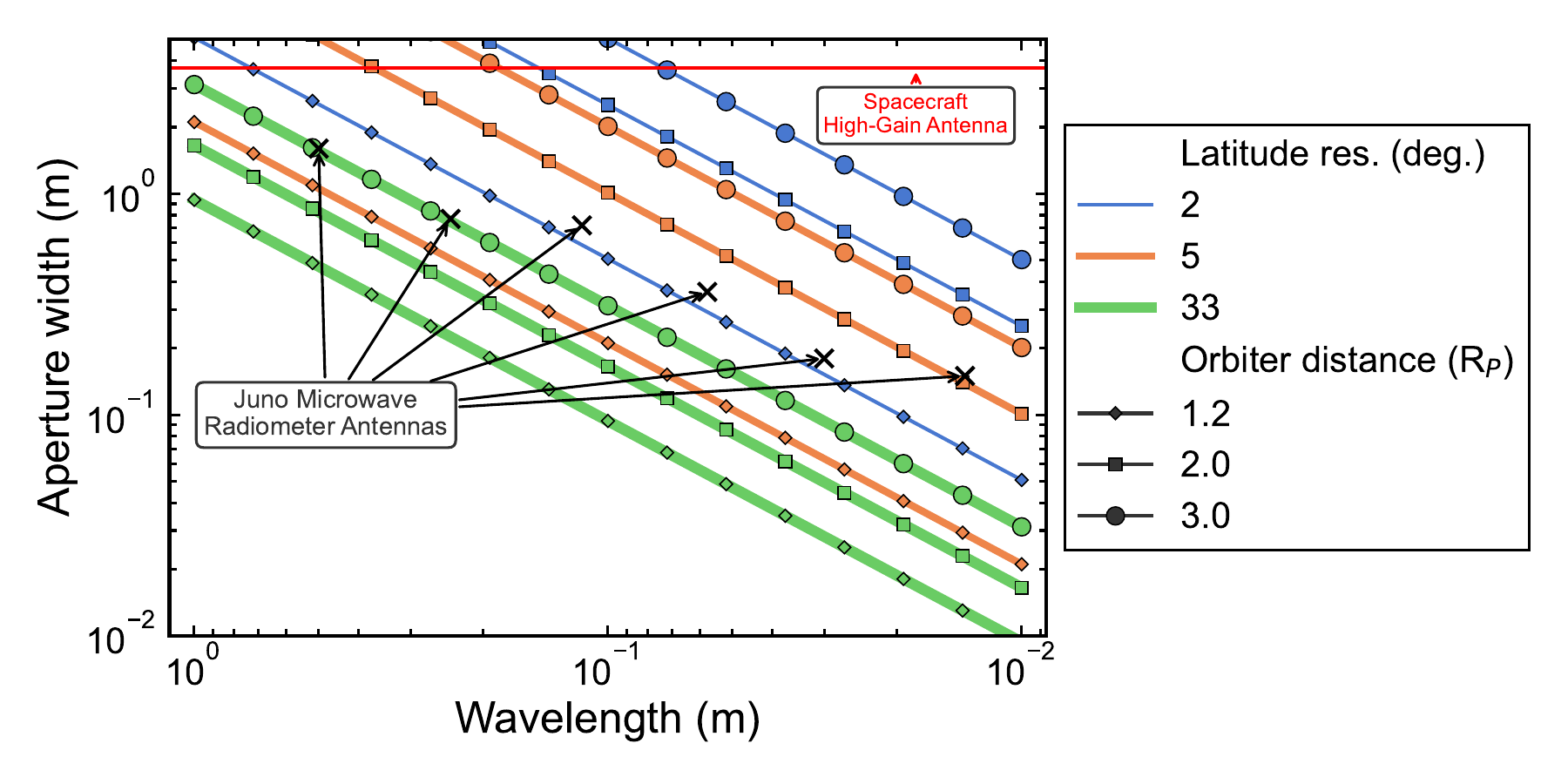} 
	\includegraphics[width=0.8\linewidth]{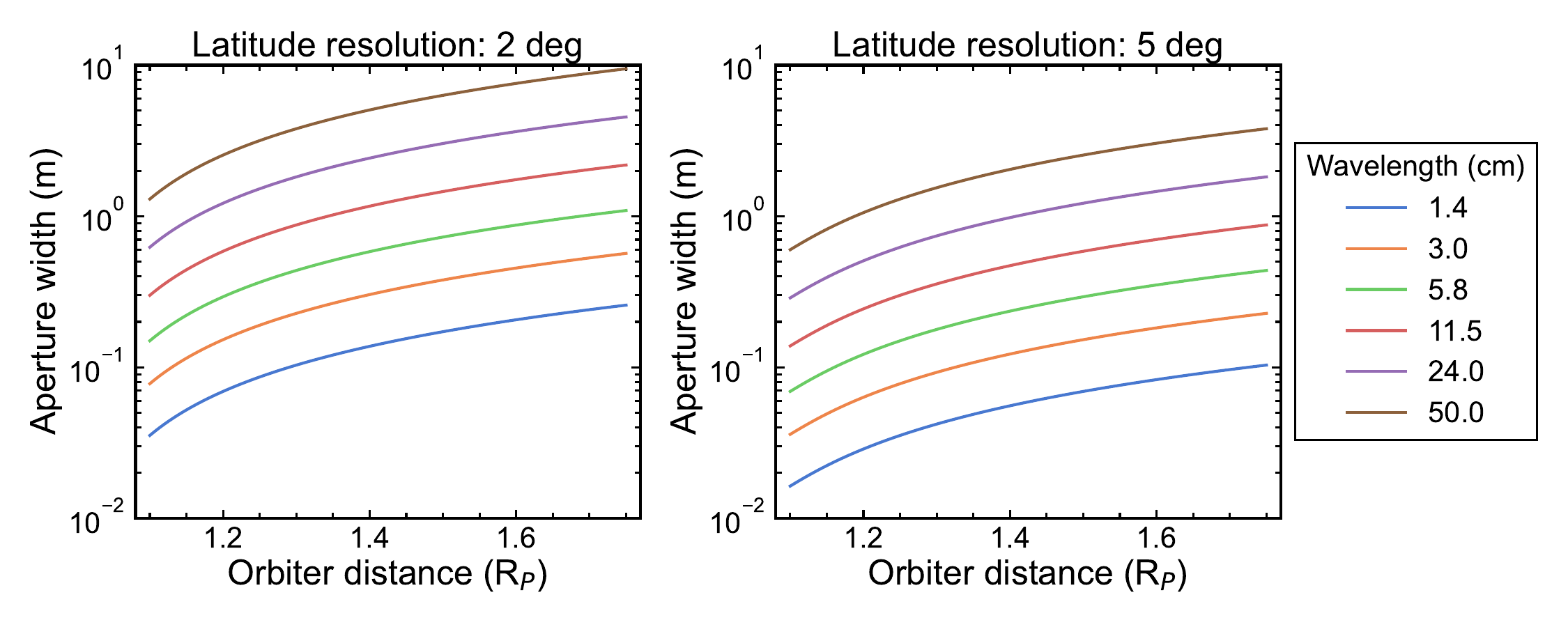}
	\includegraphics[width=0.66\linewidth]{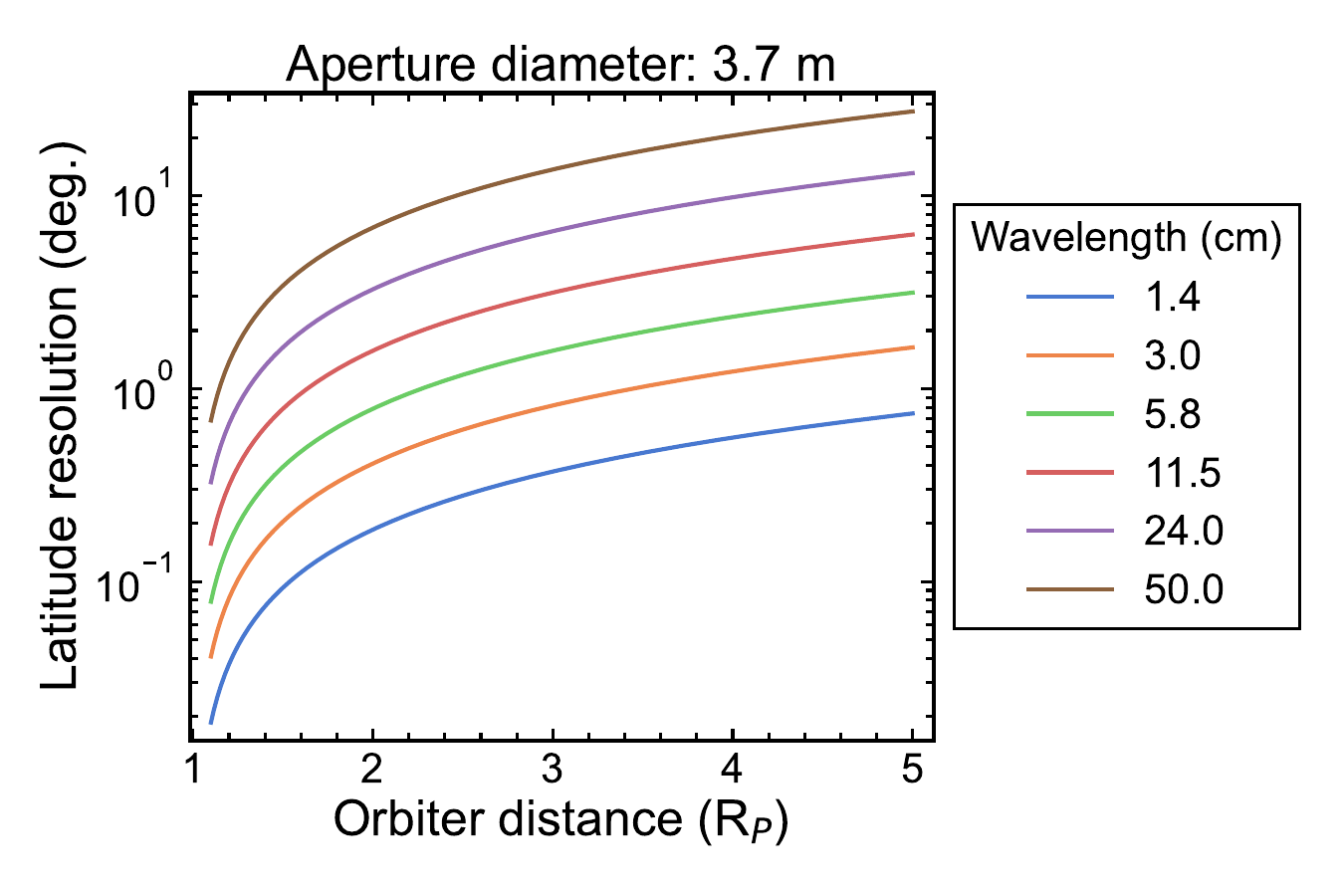}
   \caption{Aperture sizing calculations for microwave remote sensing of giant planets. (Top) Aperture widths required to achieve 2, 5, or 33 deg. on-planet latitude resolution for different orbiter distances as a function of wavelength. The sizes of the Juno MWR antennas and a 3.7 m HGA are included. (Middle) Aperture widths required to achieve 2 and 5 degree latitude resolution for different orbiter distances. Different curves correspond to the Juno MWR channel wavelengths. (Bottom) Achievable latitude resolution for a 3.7 m HGA radiometer over a range of orbiter distances for Juno MWR channel wavelengths.} 
\label{fig:apsize}
\end{figure}

MWR's are diffraction-limited systems achieving angular resolution $\theta \propto \lambda/D$ for a given aperture diameter $D$. For long-wavelength MWR channels, the necessary antenna size to obtain high spatial resolution can drive the mass requirements of the instrument. Prior missions have employed two approaches in the design of MWR apertures: using independent antenna systems (e.g. Juno MWR) or placing a feed horn near the focus of the spacecraft telecommunications high-gain antenna (HGA, e.g. Cassini RADAR). HGAs for prior outer planet missions range between 2-4 m in diameter; we use as a representative number the 3.7 meter diameter for the Voyager and Magellan mission HGAs. We provide in Figure \ref{fig:apsize} some calculations which relate the size of the aperture and the distance of the orbiter from the planet to the resulting on-planet angular resolution as a function of wavelength. 

Spatial resolution requirements are dependent on the goals of the measurement. While dividing the planet into thirds can potentially be sufficient for studying equator/pole contrasts, higher resolution is necessary to study atmospheric dynamics and wave patterns. MWR measurements generally prefer closer orbital distances to maximize spatial resolution, although it may be necessary to operate the instrument from farther away to balance the measurement requirements of other instruments or to avoid orbital debris (e.g. ring particles). While the study of atmospheric dynamics favors high resolution, it may be necessary to make marginal sacrifices in aperture size to ensure that the mass of the instrument, particularly the longer wavelength channels, does not become excessive. Such was the case for the design constraints of the Juno mission. Additionally, use of the HGA as the radiometer aperture is particularly attractive if mission constraints prevent close periapses. There are a few complications to this approach, including the need to co-boresight the instrument with telecommunications receivers (also used for gravity and occultation measurements) and the varying spatial resolution with wavelength that results from use of a fixed aperture size.      
 
\section{Conclusion} 
As demonstrated from the Juno mission, multi-channel MWR's are critical tools for studying giant planet atmospheres beneath the visible cloud tops. Future missions to Saturn and the ice giants would benefit greatly from their inclusion within a comprehensive payload architecture. When properly calibrated, these instruments can provide scientifically useful estimates of the deep abundances of volatile gases as well as insight into how the deep atmosphere transfers heat and mass. Our discussion extends pre-Juno assessments of Jupiter microwave remote sensing by \cite{Janssen2005} and \cite{DePater2005} to Saturn and Uranus. In this article, we provided a summary of MWR measurement processing and calibration relationships suitable for systems engineering assessments, and detailed how atmospheric brightness temperatures can be computed from microwave radiative transfer and equilibrium condensation models. We applied these models to illustrate the sensitivity of MWR measurements to differing deep abundances of N, S, and O for Saturn and Uranus.  We computed the standard information-content metrics and estimated the achievable vertical resolutions from different instrument configurations. Finally, we provided a brief discussion of aperture sizing and orbital distance. 

Salient points from our discussion include: 

\begin{itemize}
\item MWR observations provide a unique avenue to constrain abundances, temperatures, and circulation in the deep atmosphere of giant planets. Single point measurements with atmospheric probes cannot achieve important science objectives along these lines. 
\item Juno's MWR measurements have re-written our understanding of giant planet atmospheres and are driving community efforts towards the development of more realistic models.
\item Simple forward models remain useful in guiding future instrument design.  An MWR with $\sim$5-10 channels that observes locations from $\sim$3 look angles should be considered an effective starting point for point designs, subject to refinement when detailed science objectives are selected.
\item Ground-based observatories, current and planned, cannot achieve the necessary sensitivity to ensure precise recovery of deep atmosphere abundances at Uranus or Neptune. While next-generation arrays may achieve commensurate imaging sensitivity at shorter wavelengths for observations of Saturn, their absolute calibration and viewing geometries will remain limited.
\item MWR instruments can make high-precision measurements, but the achievable spatial resolution is generally poorer than for shorter-wavelength cameras and spectrometers.  To observe features with spatial scales much smaller than the atmospheric Rossby deformation radius, a spacecraft must either carry large antennas or pass close to its target.
\item Long-wavelength observations ($>$ 50 cm) from orbiters should be considered essential for any attempt to constrain O/H. Even un-resolved measurements (using electrically small antennas) would be of great value, although moderate-resolution imaging is greatly preferred. 
\end{itemize}

Beyond our discussion of atmospheric measurements, MWR observations have also found interesting applications in the study of giant-planet moons. For example, \cite{Brown2023a, Zhang2023a, levin_europas_2025} have used Juno MWR data to constrain the ice shell depth for Ganymede and Europa, and results from MWR observations of Io have recently been published \cite{brownIoSubSurfaceTemperature2026}. Finally, while our discussion is based around an MWR's capability to measure what we expect, some of Juno MWR's greatest successes were in obtaining measurements that were unexpected prior to orbital insertion. Similar surprises await us further in the outer solar system.

\section*{Acknowledgments}
We thank two anonymous reviewers for their assistance in clarifying our discussion. A.A. acknowledges discussions with members of the Juno MWR science team, particularly S. Brown, S. Misra, S. Levin, and S. Bolton, which have influenced the organization of this article. The research described in this paper was carried out at the Jet Propulsion Laboratory, California Institute of Technology, under a contract with the National Aeronautics and Space Administration (80NM0018D0004) and sponsored by the JPL Research and Technology Development Fund. Copyright 2026. All Rights Reserved.





\appendix
\section{The meter-wavelength spectra of Saturn and Uranus} \label{app:A}

\begin{figure}[htbp]
   \centering
   \includegraphics[width=0.45\linewidth]{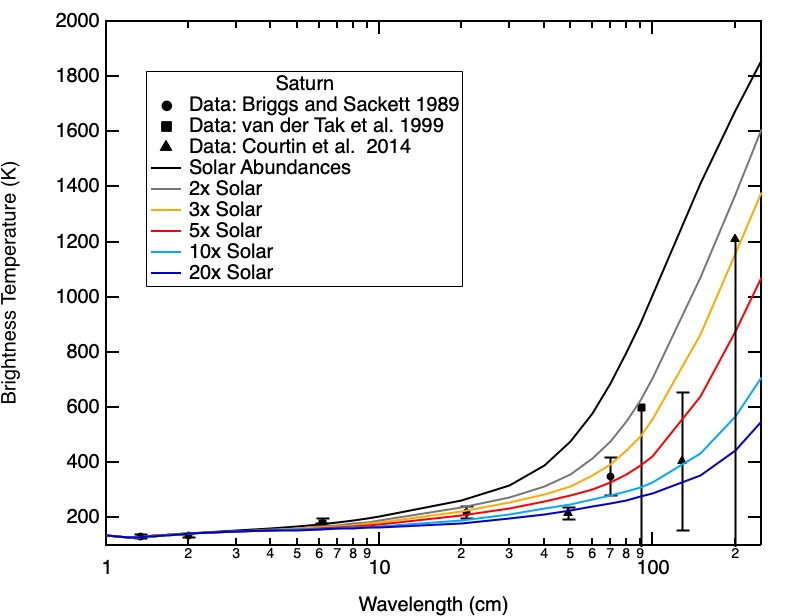} 
    \includegraphics[width=0.45\linewidth]{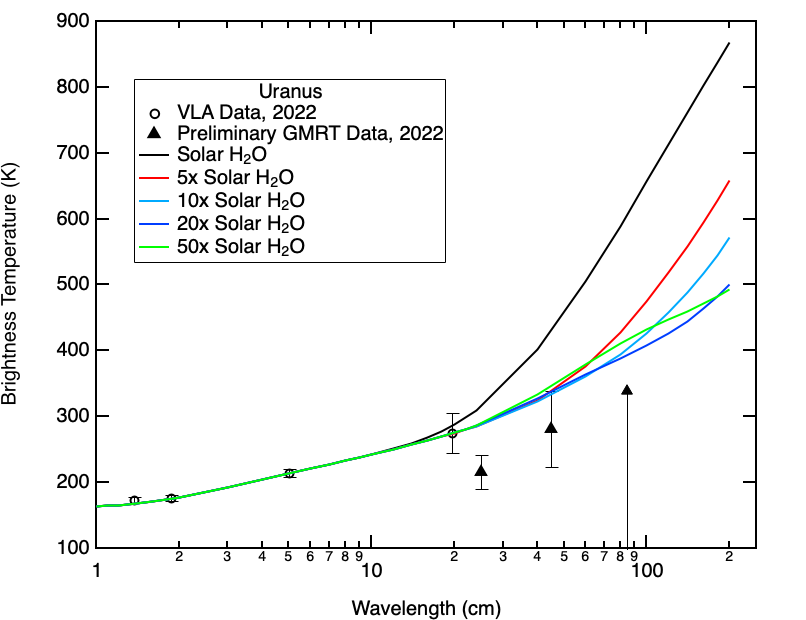}
   \caption{Ground-based centimeter-to-meter wavelength spectra for Saturn and Uranus, including observations from VLA, Arecibo, and GMRT. For Saturn, if all volatiles are equally enriched, the data suggest the H$_2$O abundance is 3-10$\times$ solar.  For Uranus, atmospheric water is currently unconstrained. In both plots, data points with one-sided error bars are non-detection upper limits. Models shown assume vapor equilibrium conditions and ECCM clouds (see Figure 1). For Saturn, the models all have volatile abundances enriched equally relative to solar, with NH$_3$ and H$_2$O being the species that most influence the spectrum. For Uranus, where volatiles appear to have drastically different abundances from one another (e.g. \cite{Akins2023}), all the models shown have NH$_3$ and H$_2$S abundances selected to fit wavelengths of 20 cm and shorter, and only variations in H$_2$O are shown.}
   \label{fig:sat_ura_meter_spec}
\end{figure}

While meter-wavelength thermal emission is overpowered by synchrotron emission at Jupiter, the same constraint does not apply for Saturn or the ice giants. Rather, the largest barriers to accurate measurement of their atmospheric brightness temperatures at present are their distances and the capabilities of ground-based observatories. Figure \ref{fig:sat_ura_meter_spec} summarizes the available constraints on the meter-wavelength brightness temperature of Saturn and Uranus. At these wavelengths, the combination of decreasing source flux and increasing background confusion results in disk brightness-temperature uncertainties dominated by observation sensitivity and calibration rather than the order 5\% uncertainty in the flux density scale \citep{Perley2017}. Observations of Saturn suggest that O/H is 3-10 $\times$ solar. Observations of Uranus were obtained by us in 2022 using the Giant Meterwave Radio Telescope (GMRT). While they demonstrate detection at wavelengths longer than 20 cm, transfer of the flux scale from observations of 3C48 suffered from direction-dependent errors, resulting in significant uncertainties in the derived brightness temperatures. Unlike the Saturn case, the Uranus long-wavelength data have insufficient accuracy for distinguishing among different models of the deep water abundance. Further GMRT observations were obtained in 2024, including a test for the detectability of Neptune, but these data have yet to be analyzed. These results, and the calculations in Section \ref{sec:calcs}, illustrate the significant improvement which would be afforded by orbital measurements. 

\bibliography{library}{}
\bibliographystyle{aasjournal}



\end{document}